\documentclass[aps, twocolumn]{revtex4-2}

\usepackage{epsfig}
\usepackage{graphicx}
\usepackage{amsmath,amssymb,amsfonts}
\usepackage{array}
\usepackage{url}
\usepackage[usenames,dvipsnames]{color}
\usepackage[colorlinks=true,linkcolor=magenta, citecolor = blue]{hyperref}
\usepackage{multirow}
\usepackage{float}
\usepackage{lineno}
\usepackage{xspace}
\usepackage{ulem}
\usepackage{lipsum} 
\usepackage{bigints}
\begin{document}

\title{Structural Properties of Magnetized Neutron Stars under $f(R,T)$ Gravity Framework}
\author{Charul Rathod}
\email{charulrathod1813@gmail.com}
\author{M. Mishra}
\affiliation{Department of Physics, Birla Institute of Technology and Science, Pilani Campus, Rajasthan, India}
\author{Prasanta Kumar Das}
\affiliation{Department of Physics, Birla Institute of Technology and Science, K. K. Birla Goa Campus, NH-17B, Zuarinagar, Sancoale, Goa 403726, India}

\begin{abstract}
The current work investigates the structural properties of neutron stars in the presence of a strong magnetic field within the framework of $f(R, T)$ modified gravity, where the matter-geometry coupling leads to deviations from general relativity at high matter densities. We present here the mass-radius sequences, as well as the mass and pressure distributions for various values of the modified gravity parameter and the central magnetic field. The modified Tolman-Oppenheimer-Volkoff equations are numerically solved using isotropic equations of state, specifically the APR, FPS, and SLy models. Comparing the corresponding results in the context of general relativity suggests that more negative values of the modified gravity parameter result in higher maximum gravitational masses. In contrast, strong central magnetic fields of up to $10^{18}$~Gauss cause only a slight decrease in maximum mass without disrupting spherical symmetry. Our findings are in agreement with the observed data from GW170817, PSR and NICER.
\end{abstract}
\maketitle
% \begin{keyword}
% Neutron Star \sep Equation of States \sep Magnetic Field \sep Modified Gravity \sep TOV equations
% \end{keyword}

\section{Introduction}
\label{intro}
Neutron stars (NSs) are the densest and most extreme astrophysical objects among other celestial objects in  nature~\cite{lattimer2004physics}. They are formed from the gravitational collapse of massive stellar cores after supernova explosions, leaving 
remnants with radii of $10$--$20$~km and typical masses of $1.4$--$2\, M_\odot$.
These compact objects also host surface magnetic fields up to $10^{13}$--$10^{14}$~Gauss, producing an environment of extreme gravity, pressure, and magnetic field~\cite{dexheimer2017magnetic}. Neutron stars, due to their extreme conditions, act as natural laboratories for studying matter under high density and gravity~\cite{nattila2022fundamental}. This makes them an ideal environment for exploring the interactions between modified gravity, dense matter physics, and ultra-strong magnetic fields~\cite{hehl1997alternative}.
They provide crucial tests for both nuclear and gravitational physics~\cite{kunz2023neutron}--in regimes unattainable in terrestrial experiments.\\

\noindent The structure of a neutron star is traditionally described within  General Relativity (GR)~\cite{einstein1915field}. The Tolman-Oppenheimer-Volkoff (TOV) equations, derived from Einstein’s field equations, establish hydrostatic equilibrium between the outward internal pressure and the gravitational inward attraction. They yield the mass--radius ($M-R$) relation when these are solved employing an appropriate EoS. 
\noindent Although GR has been remarkably successful, several large-scale observations – such as galaxy rotation curves, gravitational lensing, and cosmic matter distribution – suggest that either some invisible form of matter, commonly known as dark matter, exists or gravity deviates from GR in certain regimes. 
Two broad approaches attempt to explain these anomalies: 
(i) \textit{Dark-matter models}, which introduce non-luminous or non-baryonic matter interacting only through gravitational interaction. 
(ii) \textit{Modified-gravity theories}, which generalize GR itself through additional curvature or matter-coupling terms. 
Notable examples of modified gravity models include $f(R)$ and $f(R, T)$ gravity and Modified Newtonian Dynamics (MOND)~\cite{sotiriou2010f,de2010f,bauer2022spherical,alex2020covariant}.\\

\noindent As a minimal extension of GR, \(f(R)\) gravity promotes the Ricci scalar to a function \(f(R)\), effectively introducing a curvature–induced scalar degree of freedom. It modifies the stellar structure equations and can shift the mass–radius relation for suitable choices (e.g., \(f(R)=R+\alpha R^2\)) while remaining compatible with weak-field tests~\cite{sotiriou2010f,de2010f,nobleson2022comparison}. A natural continuation is to allow an explicit dependence on the matter sector via the trace \(T\), leading to \(f(R, T)\) gravity~\cite{harko2011f}. This matter–geometry coupling alters the effective equilibrium by introducing density- and pressure–dependent terms, thereby relaxing the standard conservation law for \(T^{\mu\nu}\). On contrary, \(f(R, T)\) inherits the curvature modifications of \(f(R)\) and augments them with trace–coupled effects, providing a controlled framework to assess how microphysics (encoded in \(T\)) can feed back on gravity at high-density environments such as neutron stars.\\

\noindent Earlier studies of neutron stars in \(f(R, T)\) gravity have primarily examined hydrostatic equilibrium, stability, and oscillations for various functional forms of \(f(R,T)\)~\cite{mahapatra2024neutron,carvalho2017stellar,lobato2020neutron}. However, these works lack a magnetic field when evaluating the mass-radius relation within the modified-gravity framework.

Conversely, various studies on magnetized neutron stars within the framework of general relativity have explored how strong magnetic fields influence the equation of state, pressure anisotropy, and stellar structure. However, these studies have focused primarily on these factors without incorporating the effects of gravity itself.~\cite{ferrer2010equation,ferrer2022hadron,dexheimer2017magnetic,chatterjee2019magnetic}. 
Thus, a thorough research work needs to be systematically done on how a matter-geometry coupling term of the form $f(R, T)=R+2\lambda\kappa T$, together with strong central magnetic fields ($B_c\!\sim\!10^{18}$~Gauss), influences the mass-radius relation and internal pressure profiles of neutron stars for realistic EoSs.\\

\noindent While the current work is based on established theoretical foundations; namely the modified Tolman–Oppenheimer–Volkoff (TOV) framework in $f(R, T)$ gravity and the treatment of strong magnetic fields in neutron star structure; its novelty lies in the combined and self-consistent analysis of these effects for multiple realistic equations of state (APR, FPS, and SLy). In other words, the present study extends the existing work by exploring the interplay between modified gravity and strong-field magnetic field effects at the structural properties, providing a new perspective on how alternative gravity may manifest on the equilibrium properties of magnetized compact stars.\\

\noindent We mainly focus here on \textit{modified gravity effects on magnetized neutron stars} employing the minimal matter-geometry coupling models in which $f(R,T)$ is expressed as:
$f(R,T) = R + 2\lambda \kappa T$, where $\kappa = 8\pi G / c^4$ and $\lambda$ quantify the coupling strength. This form isolates the impact of the trace term $T$ while retaining second-order field equations. The corresponding modified TOV equations~\cite{carvalho2017stellar} alter the pressure and density profiles, affecting the predicted stellar mass and radius. 
Furthermore, magnetic fields have a strong influence on the structure of the NS and the stiffness of the nuclear equation of state (EoS). The current work is devoted to the study of $M-R$ curves for NSs, also called magnetars, exhibiting internal magnetic fields up to $10^{15}$--$10^{18}$~Gauss~\cite{chatterjee2019magnetic,kaspi2017magnetars}. If a central magnetic field exceeds $10^{18}$~Gauss, it leads to a pressure anisotropy~\cite{ferrer2010equation,ferrer2022hadron} and further coupling between magnetic and gravitational effects. Here, we propose exploring magnetars with central magnetic fields up to  $10^{18}$ Gauss. Therefore, considering isotropic internal pressure within NSs appears to be a reasonable approximation. \\  

\noindent The paper is organized as follows: Section~\ref{form} describes the theoretical framework, including the equations of state, the magnetic field model, the modified field equations, and the derivation of the TOV system for $f(R, T) = R + 2\lambda\kappa T$. 
Section~\ref{result} discusses the results and analysis of the mass-radius relations, highlighting the effects of magnetic fields and modified gravity on the stellar properties, such as mass and pressure profiles. 
It also compares our theoretical predictions with the available observational constraints on the $M$--$R$ sequences obtained from PSR, NICER observations. Finally, Section~\ref{conc} summarizes the main conclusions and outlines possible directions for future work.

\section{Formalism}
\label{form}
Although certain aspects of the formalism used in the present study are similar to those employed in earlier works~\cite{chatterjee2019magnetic,yadav2024thermal,mahapatra2024neutron}, we briefly outline the key equations and assumptions here to make the discussion as self-contained as possible. We begin our formalism by specifying the three realistic equations of state (EoSs) considered in this work; APR, FPS, and SLy, which are used to close the system of Tolman–Oppenheimer–Volkoff (TOV) equations and determine the equilibrium structure of neutron stars in the $f(R, T)$ gravity framework.\\

\noindent In the present analysis, the TOV equations are formulated within the minimal matter-geometry coupling model $f(R, T)=R+2\lambda\kappa T$, where $\lambda$ is the coupling parameter and $\kappa=8\pi G/c^{4}$. To account for the influence of strong magnetic fields, the magnetic energy density is included in the total energy-momentum tensor. 
This treatment enables the magnetic contribution, expressed as $B^{2}/8\pi$, to directly modify the pressure and energy density terms that enter the TOV equations. 
Consequently, the equilibrium configuration reflects both the modified gravity effects (arising from the $T$-dependent coupling) and the magnetic field contributions, leading to a coupled system that governs the stellar mass-radius and pressure profiles in the presence of strong  central magnetic fields ($B_c\!\sim\!10^{18}$~Gauss).

\subsection{Equations of State}
\noindent {\textit{Akmal-Pandharipande-Ravenhall (APR) EoS}}: The APR EoS~\cite{akmal1998equation} uses isospin asymmetry~\cite{schneider2019akmal} and baryon density to model the interaction potential. The transition from the high-density phase (HDP) to the low-density phase (LDP) is captured. An important aspect of comprehending the evolution of neutron stars is this phase change, which notably causes the star to shrink more quickly.\\

\noindent {\textit{Friedman–Pandharipande–Skyrme (FPS) EoS}}: The inner crust and liquid core of the neutron star are described cohesively by the FPS EoS~\cite{flowers1976neutrino}. The transition from crust to liquid core occurs at a higher density, $\rho_{\text{edge}} = 1.6 \times 10^{14}$ gm/cm$^3$. It is preceded by a sequence of phase transitions between various nuclear shapes.\\ 

%. When considering an alternate effective nuclear Hamiltonian, a crucial factor in neutron star matter modeling, the FPS EoS is especially appropriate.~\cite{flowers1976neutrino.\\

\noindent {\textit{Skyrme–Lyon (SLy) EoS}}: The SLy effective nuclear interaction model proposes that nuclei within the neutron star matter maintain a spherical shape down to the bottom of the inner crust. Once the density reaches $(\rho = 1.3 \times 10^{14} \, \text{gm/cm}^3 )$, a uniform neutron-proton-electron (npe) plasma is formed. The cooling behavior of the star is well described by this EoS, particularly near the minimum mass, where the liquid core and solid crust converge. \\
The SLy EoS differs from the FPS EoS in that it employs the same nuclear Hamiltonian to describe both the liquid core (uniform nuclear matter) and the inner crust (nuclear structures). In contrast to the FPS EoS, the SLy EoS is stiffer in the region of the crust-core interface. Moreover, the SLy model demonstrates a more pronounced discontinuous stiffening at the crust-core transition compared to the FPS model.\\

\noindent The interior structure and behavior of neutron stars, especially in relation to the phase transitions between the crust and core, as well as the thermodynamic characteristics of matter at extremely high density, are described differently by these three EoSs.

\subsection{TOV Equations for GR Case}
\label{subsec:TOV_GR}
\noindent In order to describe a static, spherically symmetric, and non-rotating compact object in hydrostatic equilibrium, we adopt the Schwarzschild-like metric.
\begin{equation}
ds^{2} = -e^{2\phi(r)}dt^{2} + e^{2\Lambda(r)}dr^{2} + r^{2}\left(d\theta^{2} + \sin^{2}\theta\,d\Phi^{2}\right),
\label{eq:metric}
\end{equation}
where $\phi(r)$ and $\Lambda(r)$ are functions of the radial coordinate $r$ representing the gravitational potential and spatial curvature, respectively.  
The metric function $\Lambda(R)$ is related to the mass function $m(R)$ as:
\begin{equation}
e^{-2\Lambda(R)} = 1 - \frac{2G\,m(R)}{R c^{2}},
\label{eq:lambda}
\end{equation}
ensuring that the solution matches the exterior Schwarzschild geometry at the stellar surface. The potential $\phi(r)$ satisfies the surface boundary condition $e^{2\phi(R)} = 1 - 2GM/(Rc^{2})$, where $M=m(R)$ is the total gravitational mass.

\noindent Using the above metric, Einstein's field equations for a perfect fluid distribution yield the classical Tolman-Oppenheimer-Volkoff (TOV) equations~\cite{tolman1939static,oppenheimer1939massive}:
\begin{equation}
\frac{dm}{dr} = 4\pi r^{2}\rho(r),
\end{equation}
\begin{equation}
\frac{d\phi}{dr} = \frac{G\left[m(r) + 4\pi r^{3}P/c^{2}\right]}{r\left(rc^{2} - 2Gm(r)\right)},
\end{equation}
\begin{equation}
\frac{dP}{dr} = -c^2\left(\rho(r) + P/c^2\right)\frac{d\phi}{dr},
\end{equation}
where $\rho(r)$ and $P(r)$ denote the local density and pressure, respectively. $m(r)$ represents the mass enclosed within the radius $r$. 
The metric potential $\phi(r)$ corresponds to the gravitational potential at each radius.

\noindent The system is closed with an appropriate equation of state (EoS) (APR, FPS and SLy), $P=P(\rho)$, describing dense nuclear matter. 
Numerical integration is carried out from the center $(r=0,\,P=P_c)$ outward to the surface $(r=R,\,P=0)$, where $R$ denotes the stellar radius and $M=m(R)$ the total gravitational mass~\cite{hartle1968slowly,schleich2009does}. This formulation provides a general relativistic description of neutron star equilibrium, yielding the mass-radius relation and internal pressure and density profiles for a particular EoS. 

%%%%%%%%%%%%%%%%%%%%%%%%%%%%%%%%%%%%%%%%%%%
\noindent In magnetized neutron stars (magnetars), strong central magnetic fields ($B_c\!\sim\!10^{17\text{--}18}\,\mathrm{Gauss}$) modify the internal energy density and pressure distributions.  
The electromagnetic stress-energy tensor adds an isotropic magnetic energy density $B^{2}/8\pi$, altering the equilibrium structure~\cite{chatterjee2019magnetic}. Here $B$ is the magnitude of the magnetic field at a radial distance $r$, i.e., $B=B(r)$.  
The TOV equations can therefore be modified as:
\begin{equation}
\frac{dm}{dr} = 4 \pi r^2 \left( \rho(r) + \frac{B^2}{8 \pi c^2} \right),
\label{eq:mass_mag}
\end{equation}
\begin{equation}
\frac{d\phi}{dr} = \frac{G \left( m(r) + 4 \pi r^3 {P/c^2} \right)}{r \left( r c^2 - 2 G m(r) \right)},
\end{equation}
\begin{equation}
\frac{dP}{dr} = -c^2 ( \rho(r) + \frac{B^2}{8 \pi c^2} +P/c^2) \left( \frac{d\phi}{dr} - \mathcal{L}(r) \right),
\label{eq:pressure_mag}
\end{equation}
where $\mathcal{L}(r)$ denotes the Lorentz-force contribution and parameterized as~\cite{dexheimer2017magnetic}:
\begin{equation}
\mathcal{L}(r) = B_{c}^{2}\,[-3.8x + 8.1x^{3} - 1.6x^{5} - 2.3x^{7}]\times 10^{-41},
\qquad x = \frac{r}{\bar{r}}.
\label{eq:Lorentz}
\end{equation}
Here, $B_{c}$ represents the central magnetic field, and $\bar{r}$ is a characteristic stellar radius, typically slightly larger than $R$.  
The radial variation of the magnetic field is prescribed by an empirical polynomial profile~\cite{dexheimer2017magnetic}:
\begin{equation}
B(r) = B_{c}\!\left[1 - 1.6x^{2} - x^{4} + 4.2x^{6} - 2.4x^{8}\right],
\label{eq:Bprofile}
\end{equation}
which smoothly decreases from a maximum at the center of NSs to a comparatively lower value at the surface.  

\noindent Although ultra-strong magnetic fields can induce pressure anisotropy ($P_{\parallel}\neq P_{\perp}$)~\cite{ferrer2010equation,ferrer2022hadron} due to extra magnetization term, we retain the isotropic approximation since the contribution of magnetization satisfies $MB/P \lesssim 0.05$ even for central magnetic field, $B_{c}\!\sim\!10^{18}\,\mathrm{Gauss}$~\cite{broderick2000equation}. 
This ensures spherical symmetry and allows for a consistent treatment using a single radial pressure, $P(r)$.  

\noindent Hence, Eqs.~\eqref{eq:mass_mag}--\eqref{eq:pressure_mag} represent the modified TOV equations due to the strong magnetic field.
%---------------------------------------------
\subsection {TOV equations in modified $f(R,T)$ gravity}
\label{subsec:fRT_TOV}
The $f(R,T)$ gravity is the extension of the $f(R)$ framework~\cite{nobleson2022comparison}. It generalizes the Einstein-Hilbert action by incorporating an explicit dependence on the trace of the energy-momentum tensor \(T\). Hereafter, we employ geometrized units with \( G = c = 1 \).\\
The action is expressed as~\cite{harko2011f}:
\begin{equation}
S = \frac{1}{2\kappa}\int d^4x\,\sqrt{-g}\,f(R,T) + S_m(g_{\mu\nu},\Psi),
\label{eq:action_fRT}
\end{equation}
where \(\kappa = 8\pi \), \(R\) is the Ricci scalar, and \(S_m\) is the action corresponding to matter.  
Variation with respect to the metric yields the modified field equations~\cite{sotiriou2010f,de2010f}:
\begin{equation}
f_R R_{\mu\nu} - \tfrac{1}{2}f g_{\mu\nu} + (g_{\mu\nu}\Box - \nabla_\mu\nabla_\nu)f_R
= \kappa T_{\mu\nu} - f_T (T_{\mu\nu} + \Theta_{\mu\nu}),
\label{eq:field_general}
\end{equation}
where \(f_R=\partial f/\partial R\), \(f_T=\partial f/\partial T\), and 
\(\Theta_{\mu\nu}=g^{\alpha\beta}\delta T_{\alpha\beta}/\delta g^{\mu\nu}\).\\

For the model \(f(R,T)=R+2\lambda\kappa T\) we have \(f_R=1\) and \(f_T=2\lambda\kappa\) (with \(\lambda\) the coupling parameter), which leads to~\cite{pretel2021neutron,mahapatra2024neutron}
\begin{equation}
R_{\mu\nu}-\tfrac{1}{2}R g_{\mu\nu}
= \kappa T_{\mu\nu} - 2\lambda\kappa\,(T_{\mu\nu}+\Theta_{\mu\nu})
+ \lambda\kappa\, T\, g_{\mu\nu}.
\label{eq:field_RplusT}
\end{equation}
Since \(f_R=1\), the field equations remain second order and no extra scalar (as in \(f(R)\)) is introduced. The deviations from GR therefore enter solely through the algebraic terms \(f_T=2\lambda\kappa\) and the modified conservation law.\\

Taking the trace of Eq.~\eqref{eq:field_RplusT} gives:
\begin{equation}
R = -\kappa T - 6\lambda\kappa T - 8\lambda\kappa\rho,
\label{eq:trace_fRT}
\end{equation}
showing that curvature is algebraically linked to local matter variables (\(\rho, P\)).  
Thus, \(f(R, T)\) gravity remains a second-order theory that can be solved as an initial value problem without introducing an extra scalar field.\\

\noindent 
Projecting the modified field equations of $f(R, T)$ gravity onto the $tt$ and $rr$ components of the spherically symmetric metric~\eqref{eq:metric} yields the modified structure equations for stellar equilibrium.
From the mixed $tt$ components, where \(G^{t}{}_{t} = -2m'(r)/r^{2}\), one obtains the differential equation for the enclosed mass~\cite{moraes2016stellar,carvalho2017stellar}:
\begin{equation}
\frac{dm}{dr} = \frac{\kappa}{2}\,r^{2}\,\rho - \frac{r^{2}}{4}\,h(T),
\label{eq:dm_general}
\end{equation}
where \(h(T)\) represents the functional dependence of $f(R,T)$ on the trace \(T\) due to the geometry–matter coupling.

\noindent 
For the linear model \(f(R,T)=R+2\lambda\kappa T\), one has \(h(T)=2\lambda\kappa T\), and Eq.~\eqref{eq:dm_general} simplifies as:
\begin{equation}
\frac{dm}{dr} = 4\pi r^{2}\rho - \frac{r^{2}}{4}\,h(T),
\qquad
T = -\rho + 3P, \quad \kappa = 8\pi.
\label{eq:dm_linear}
\end{equation}

\noindent 
Combining the momentum conservation equation (\(\nabla_{\mu}T^{\mu\nu}=0\)) with the $rr$ and $tt$ components of the modified field equations leads to the generalized TOV system for $f(R, T)=R+2\lambda\kappa T$ gravity:
\begin{equation}
\begin{aligned}
\frac{d\phi}{dr} &=
\frac{
m + 4\pi r^{3} P
+ \dfrac{\lambda\kappa}{2}\,r^{3}\left[T + 2(\rho + P)\right]
}{
r(r - 2m)
}, \\[6pt]
\end{aligned}
\end{equation}

\begin{equation}
\begin{aligned}
\frac{dP}{dr} &=
-(\rho + P)\,\frac{d\phi}{dr}.
\end{aligned}
\label{eq:TOV_fRT}
\end{equation}

\noindent 
In the limit \(\lambda \rightarrow 0\), Eqs.~\eqref{eq:dm_linear} and~\eqref{eq:TOV_fRT} consistently reduce to the standard TOV equations of General Relativity, thereby ensuring that the present formulation smoothly recovers the GR case when the matter–geometry coupling vanishes.

%---------------------------------------------

\subsection{TOV equations with Magnetic Field and Modified Gravity}
\label{subsec:combined}

In the presence of a strong internal magnetic field, the energy density and pressure get additional contributions, while the curvature--matter coupling depends only on the trace \(T=-\rho+3P\), of the matter energy momentum tensor, since the electromagnetic part of the energy momentum tensor is traceless.  
The resulting coupled equations are:
\begin{equation}
\frac{dm}{dr} = 4\pi r^2\!\left(\rho + \frac{B^2}{8\pi}\right) - \frac{r^2}{4}\,h(T),
\label{eq:mass_total}
\end{equation}

\begin{equation}
\frac{d\phi}{dr} =
\frac{
  m + 4\pi r^{3} P
  + \dfrac{\lambda\kappa}{2}\, r^{3}
    \bigl[\, T + 2\bigl(\rho + \dfrac{B^{2}}{8\pi} + P \bigr) \bigr]
}{
  r\,\bigl(r - 2m\bigr)
}\,,
\label{eq:phi_total}
\end{equation}

\begin{equation}
\frac{dP}{dr} = 
-\!\left(\rho + \frac{B^2}{8\pi } + {P}\right)
\!\left(\frac{d\phi}{dr} - \mathcal{L}(r)\right),
\label{eq:P_total}
\end{equation}
with \(B(r)\) and \(\mathcal{L}(r)\) given by Eqs.~\eqref{eq:Bprofile} and~\eqref{eq:Lorentz}, respectively.  
These equations are solved using the boundary conditions \(m(0)=0\), \(\rho(0)=\rho_c\), and \(P(R)=0\). The resulting solutions yield the stellar mass \(M=m(R)\) and radius \(R\) for given parameters \(\lambda\) and \(B_c\).

This unified framework consistently recovers: equations in equations 
\begin{itemize}
\item \textbf{General Relativity:} for \(\lambda \to 0, B \to 0\);
\item \textbf{Magnetized GR Star:} for \(\lambda \to 0, B \neq 0\);
\item \textbf{Non-magnetic $f(R,T)$ Star:} for \(\lambda \neq 0, B = 0\).
\end{itemize}

%---------------------------------------------
%%%%%%%%%%%%%%%%%%%%%%%%%%%%%%%%%%%%%%%%%%%%%%%%%%%%%%%%%%%%%%%%%%%%%%%%%%%%%%

\noindent The coupled equations~\eqref{eq:mass_total}–\eqref{eq:P_total} form a system of first-order differential equations for \(\{m(r),P(r),\phi(r)\}\).  
They are integrated outward using a fourth-order Runge-Kutta method with adaptive step control.
% Initial conditions at the center are expressed as:
% \[
% m(0)=0, \qquad \rho(0)=\rho_c, \qquad P(0)=P_c,
% \]
% and the integration is carried out outward until \(P(R)=0\).  
At each step, the magnetic field \(B(r)\) and Lorentz term \(L(r)\) are evaluated via Eqs.~\eqref{eq:Bprofile} and~\eqref{eq:Lorentz}.  
The final stellar radius \(R\) and total mass \(M=m(R)\) define the mass--radius relation for each equation of state (APR, FPS, SLy). Recovery of the GR limit (\(\lambda \!\to\! 0,\, B\!\to\!0\)) is used to validate the numerical accuracy.\\

\section{Results and Discussions}
\label{result}
\noindent {\textbf{\textit{Effects of Magnetic Field and EoS on the mass–radius $M-R$ Curves}}}\\

\begin{figure*}
\includegraphics[scale = 0.29]{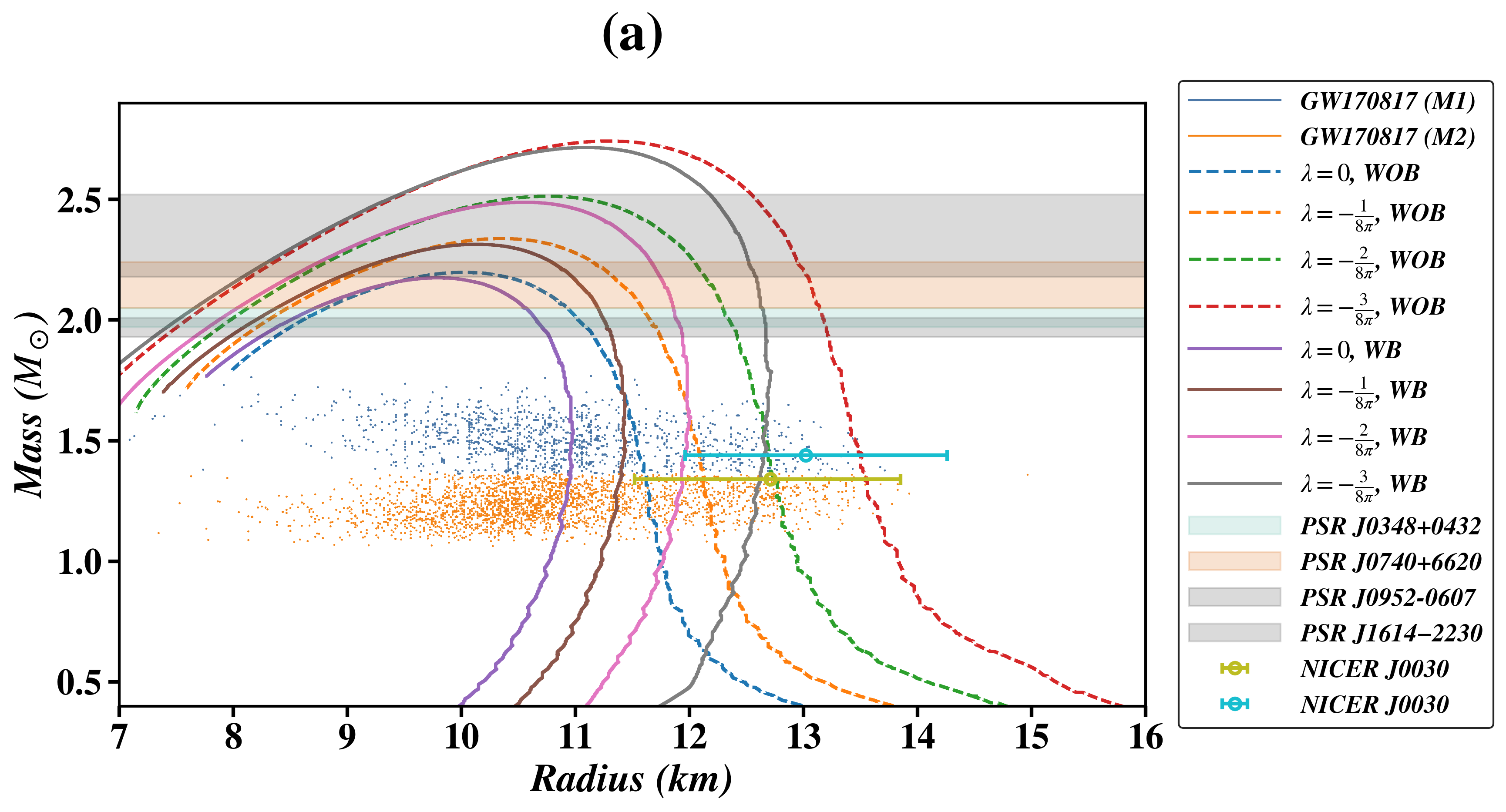}
\includegraphics[scale = 0.29]{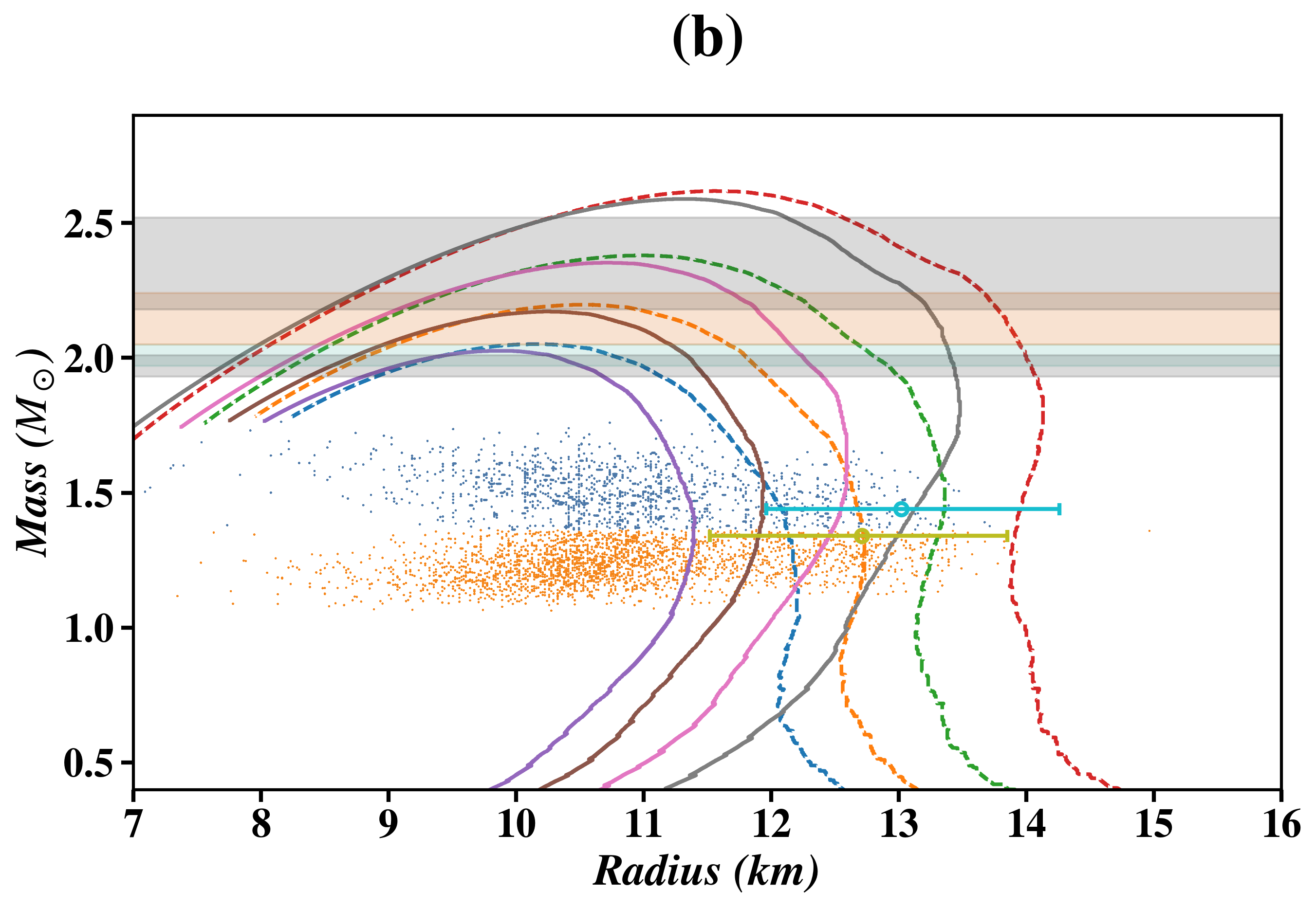}
\includegraphics[scale = 0.29]{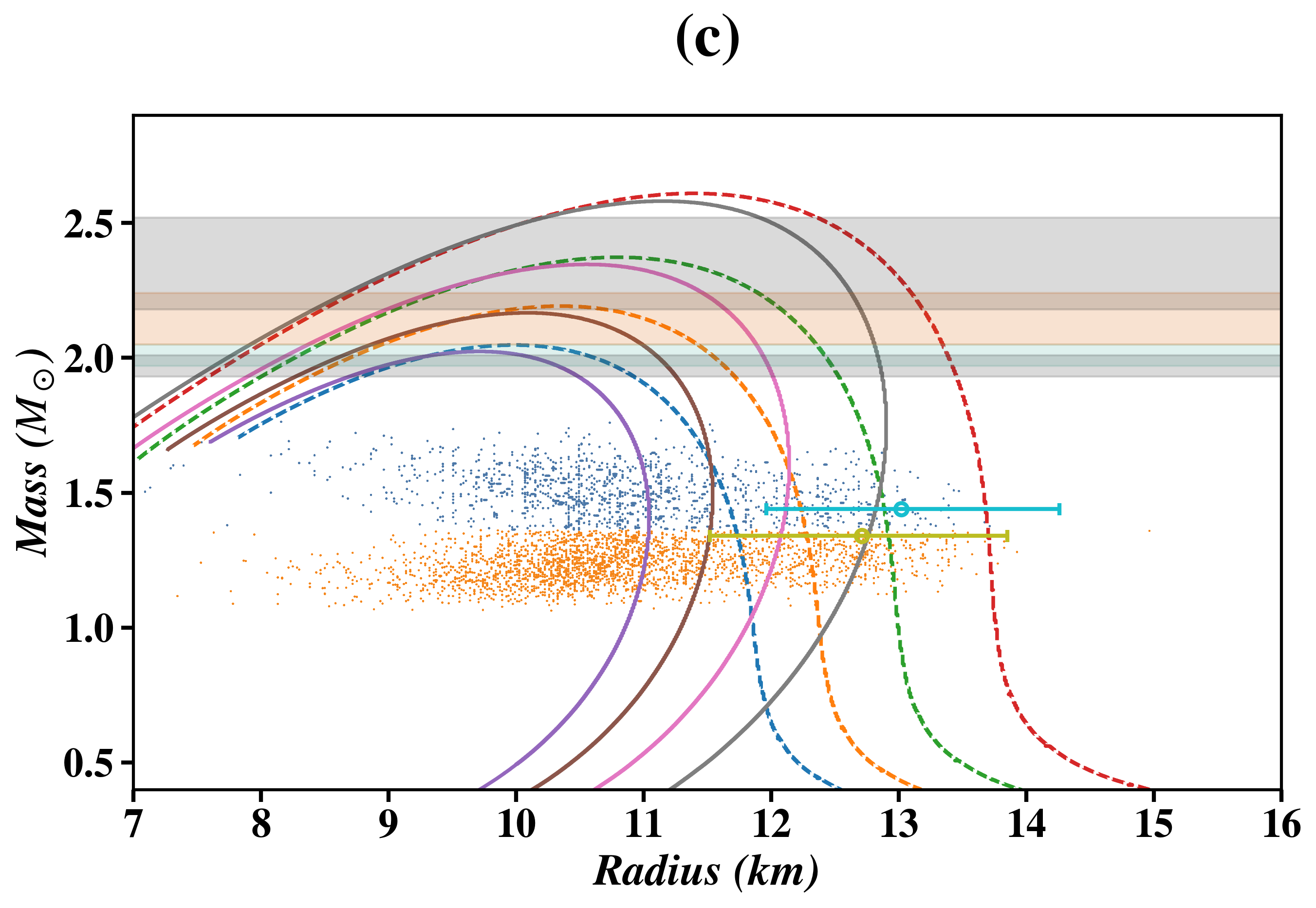}
 \caption{Mass vs radius profile for different neutron star equations of state (EoSs): (a) APR, (b) FPS and (c) SLy, both in the absence ($B_c = 0$) and presence $(B_c = 10^{18})$ Gauss of magnetic fields for the modified gravity parameter \( \lambda \)= $0$, \( -\frac{1}{8\pi} \), \( -\frac{2}{8\pi} \), and \( -\frac{3}{8\pi}\), respectively. The cases without magnetic fields are labeled as ”WoB” (Without Magnetic field), and the cases with magnetic fields are labeled as ”WB” (With Magnetic field). Overlaid observational constraints: NICER J0030+0451 ~\cite{riley2019nicer,miller2019psr}, PSR J0348+0432~\cite{antoniadis2013massive} and PSR J0740+6620~\cite{fonseca2021refined,cromartie2020relativistic}, PSR J0952-0607~\cite{romani2022psr}, PSR J1614-2230~\cite{demorest2010two} and the blue and orange cloud region is the constraints for mass-radius from the GW170817
event, which was a merger of two neutron stars with an observation in the electromagnetic and
gravitational spectrum~\cite{abbott2018gw170817}.}
 % GW170817 radius summary band at $M=1.4\,M_\odot$ ($R_{1.4}=11.9\pm1.4$ km)~\cite{abbott2018gw170817}, and horizontal mass-only bands for 
 
\label{fig:prod_wob}
\end{figure*}

\begin{table*}[ht!]
\centering
\caption{ Maximum mass and radius for different \(\lambda\) values using APR, FPS, and SLy EoSs. 
\(M_0\) and \(R_0\) correspond to \(B_c=0\), while \(M_{18}\) and \(R_{18}\) correspond to \(B_c=10^{18}\,\mathrm{Gauss}\).}
\label{tab:maxmass_radius_all}

\setlength{\tabcolsep}{7pt}
\renewcommand{\arraystretch}{1.2}

\begin{tabular}{|c|cccc|cccc|cccc|}
\hline\hline
\multirow{3}{*}{\(\lambda\)} 
& \multicolumn{4}{|c|}{\textbf{APR}} 
& \multicolumn{4}{|c|}{\textbf{FPS}} 
& \multicolumn{4}{|c|}{\textbf{SLy}} \\
\cline{2-5} \cline{6-9} \cline{10-13}
& \(M_0\) & \(R_0\) & \(M_{18}\) & \(R_{18}\)
& \(M_0\) & \(R_0\) & \(M_{18}\) & \(R_{18}\)
& \(M_0\) & \(R_0\) & \(M_{18}\) & \(R_{18}\) \\
& \([\!M_\odot]\) & [km] & \([\!M_\odot]\) & [km]
& \([\!M_\odot]\) & [km] & \([\!M_\odot]\) & [km]
& \([\!M_\odot]\) & [km] & \([\!M_\odot]\) & [km] \\
\hline
0
& 2.20 & 10.07 & 2.17 & 9.85 &
 2.05 & 10.24 & 2.03 & 9.98 &
 2.05 & 10.03 & 2.02 & 9.76 \\
\hline
\(-\tfrac{1}{8\pi}\)
& 2.34 & 10.40 & 2.31 & 10.19 &
 2.20 & 10.61 & 2.17 & 10.27 &
 2.19 & 10.42 & 2.17 & 10.16 \\
\hline
\(-\tfrac{2}{8\pi}\)
& 2.51 & 10.86 & 2.49 & 10.64 &
 2.38 & 11.03 & 2.35 & 10.75 &
 2.37 & 10.88 & 2.35 & 10.57 \\
\hline
\(-\tfrac{3}{8\pi}\)
& 2.74 & 11.37 & 2.71 & 11.17 &
 2.62 & 11.57 & 2.59 & 11.41 &
 2.61 & 11.44 & 2.58 & 11.22 \\
\hline\hline
\end{tabular}
\end{table*}

\noindent{Figures~\ref{fig:prod_wob}} depict the mass–radius, $M-R$ curves for NSs using APR, FPS, and SLy EoSs satisfying the set of TOV equations within the $f(R,T)$ model in the absence (\(B=0\)) (dashed lines) and presence of a central magnetic field \(B_c=10^{18}\,\mathrm{Gauss}\) (solid lines).

\noindent For the non-magnetized APR sequence, the largest stable gravitational mass occurs at \(\lambda=-\frac{3}{8\pi}\) (Fig.~\ref{fig:prod_wob}(a)).
% Along the equilibrium sequence parameterized by the central density \(\rho_c\), configurations are radially stable at \(dM/d\rho_c>0\). The \( dM/d\rho_c=0\) defines the maximum mass \(M_{\max}\) limit.
Increasing the magnitude of negative \(\lambda\) effectively stiffens the model and shifts the sequence toward a larger \(M_{\max}\) relative to GR (\(\lambda=0\)).\\

\noindent{\it Stability Criterion:}
\noindent Magnetized mass-radius curves exhibit a clear turning point or "S-shaped" behavior near the maximum-mass configurations. According to the standard turning-point stability criterion, radial instability occurs precisely at the maximum-mass point, where $\partial M/\partial\rho_c = 0$~\cite{hadvzic2021turning,harrison1965gravitation}. Configurations on the ascending branch ($\partial M/\partial\rho_c > 0$) are stable against radial perturbations, whereas those on the descending branch ($\partial M/\partial\rho_c < 0$) are unstable and would collapse even if a small perturbation is present~\cite{dexheimer2012hybrid,kokkotas2001radial}. Strong magnetic fields accentuate this instability limit by adding gravitational binding through their energy density, making the back-bending more pronounced compared to non-magnetized cases~\cite{bocquet1995rotating,chatterjee2015consistent,franzon2016self}. Hence, only configurations up to the turning point represent physically realizable and stable neutron stars.
\\

\noindent Including a central field \(B_c=10^{18}\,\mathrm{Gauss}\) (within the isotropic-pressure approximation) shifts the APR curves slightly to smaller radii and yields a modest decrease in \(M_{\max}\).
This behavior reflects that the magnetic energy density \(\varepsilon_B=B^2/(8\pi)\) adds to the gravitational mass but does not provide additional isotropic pressure support, thereby increasing compactness at a given \(\rho_c\).\\

% \noindent In the non-magnetized case, the maximum gravitational mass is obtained for $\lambda = -\frac{3}{8\pi}$ using APR EoS in Figure \ref{fig:prod_wob}(a). This indicates that more negative values of $\lambda$ lead to stiffer configurations and result in shifting toward the maximum mass limit compared to the GR case ($\lambda=0$). 

% \noindent For a central magnetic field, \(B_c=10^{18}\,\mathrm{Gauss}\), the figure shows the slight displacement of curves toward smaller radii and a smaller maximum mass limit for APR EoS. The reduction in radius occurs because the magnetic energy density \(\varepsilon_B = B^2/(8\pi)\) strengthens the gravitational binding of the star, which surpasses the supporting isotropic outward pressure. Consequently, a strong internal magnetic field causes the star to become more compact, resulting in a higher central density. Hence, the maximum mass, beyond which a neutron star becomes unstable, is reduced.

\noindent For $\lambda<0$, the effective gravitational coupling is reduced, equilibrium is reached at lower central density, and the $M$-$R$ sequence therefore shifts upward/rightward. As a consequence, both $R$ and $M_{\max}$ exceed as compared to the corresponding GR ($\lambda=0$) values. For $\lambda=0$, the coupling is enhanced, yielding smaller radii, lower $M_{\max}$, and a maximum attained at higher $\rho_c$.\\

\noindent For $\lambda<0$, adding $B_c=10^{18}\,\mathrm{Gauss}$ yields a $\sim$few-percent contraction in radius and decrease in $M_{\max}$, with the peak shifting to a higher $\rho_c$ (see Table~\ref{tab:maxmass_radius_all}, Fig.~\ref{fig:prod_wob}(a)).

\noindent $M-R$ sequences for FPS and SLy qualitatively follow a trend similar to the case of APR. A central magnetic field, $B_c=10^{18}\,\mathrm{Gauss}$ causes slight compactification and a modest drop in $M_{\max}$, while more negative $\lambda$ increases both $R$ and $M_{\max}$.\\

\noindent For comparison with our predicted $M-R$ values, we have plotted observational datasets namely; NICER posterior distributions for $J0030+0451$ and $J0740+6620$~\cite{riley2019nicer,miller2019psr}, and horizontal mass bands for the pulsars $PSR J0348+0432$ ($2.01\pm0.04\,M_\odot$)~\cite{antoniadis2013massive} , $PSR J0740+6620$ ($2.14^{+0.10}_{-0.09}\,M_\odot$)~\cite{fonseca2021refined,cromartie2020relativistic},PSR J0952-0607 \cite{romani2022psr},PSR J1614-2230\cite{demorest2010two} and GW170817\cite{abbott2018gw170817}. \\

It is worthwhile to mention here an additional observation that is associated with the case of a strong internal magnetic field. In this case, it is observed that for a particular value of $R$, two NSs are possible with different masses based on the two branches of the same $M-R$ curves. The strange behavior under the effect of a strong internal magnetic field is a consequence of stability criteria as discussed earlier.  \\

\begin{figure*}
\includegraphics[scale =0.29]{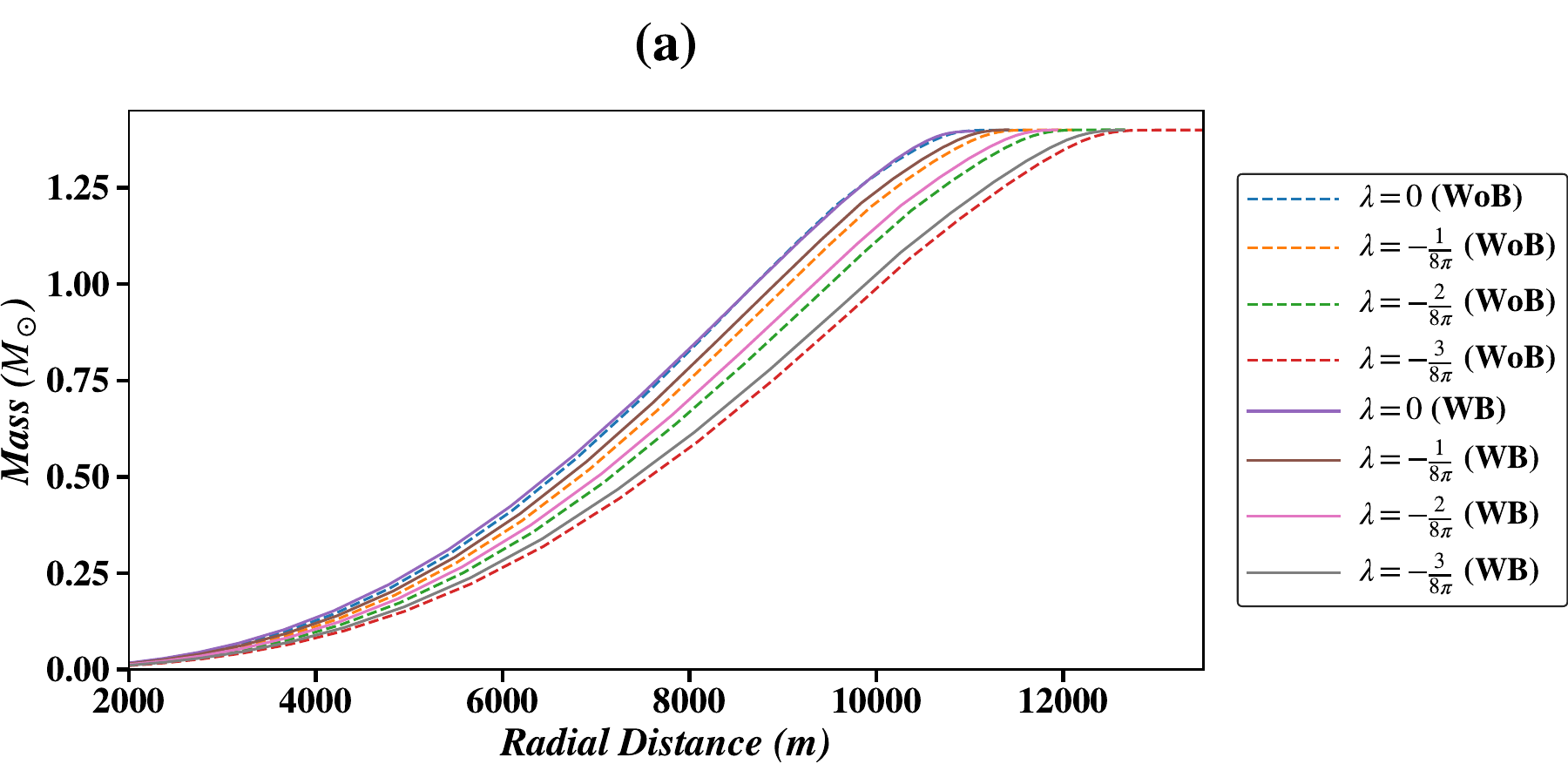}
\includegraphics[scale = 0.29]{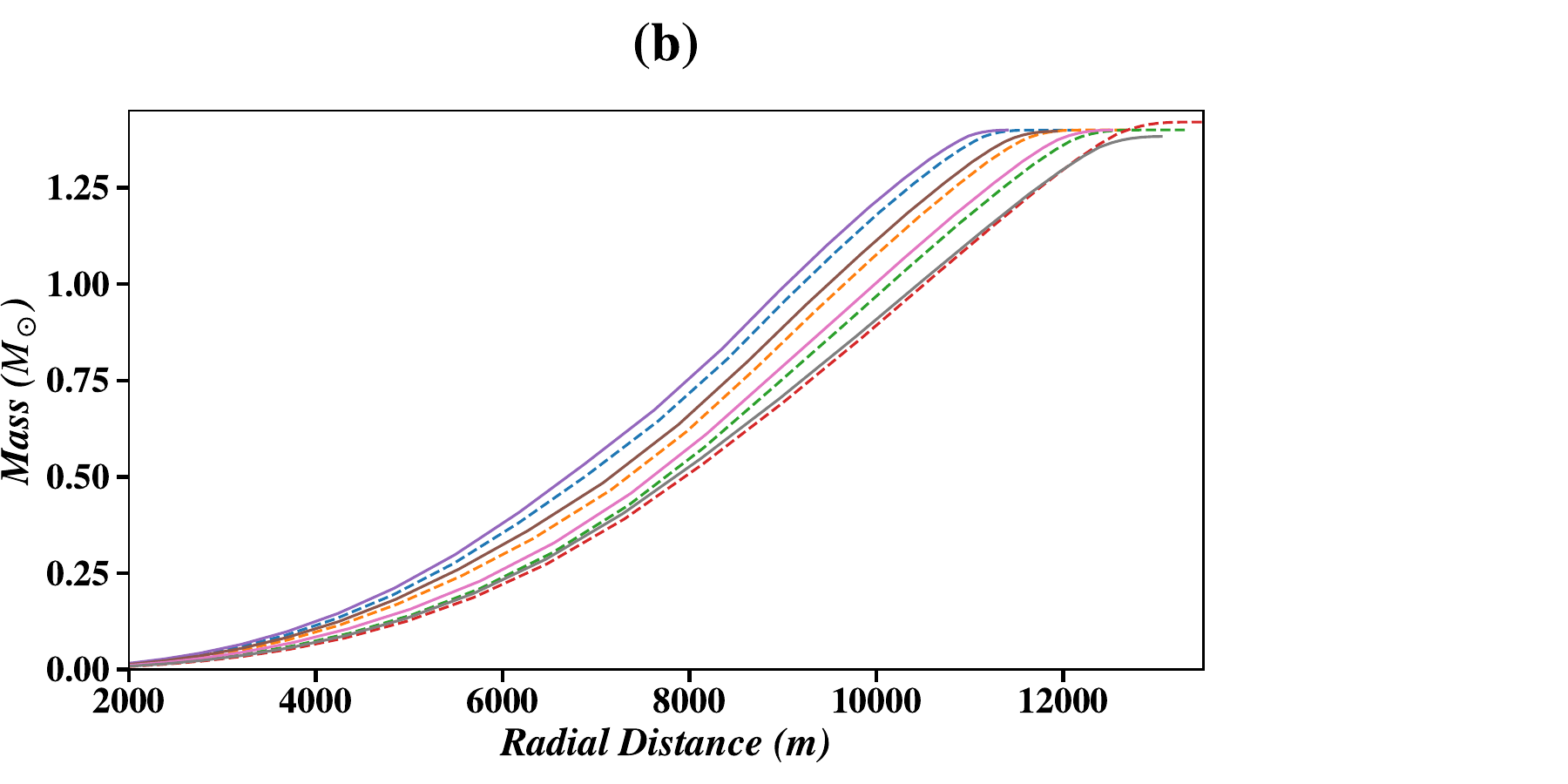}
\includegraphics[scale = 0.29]{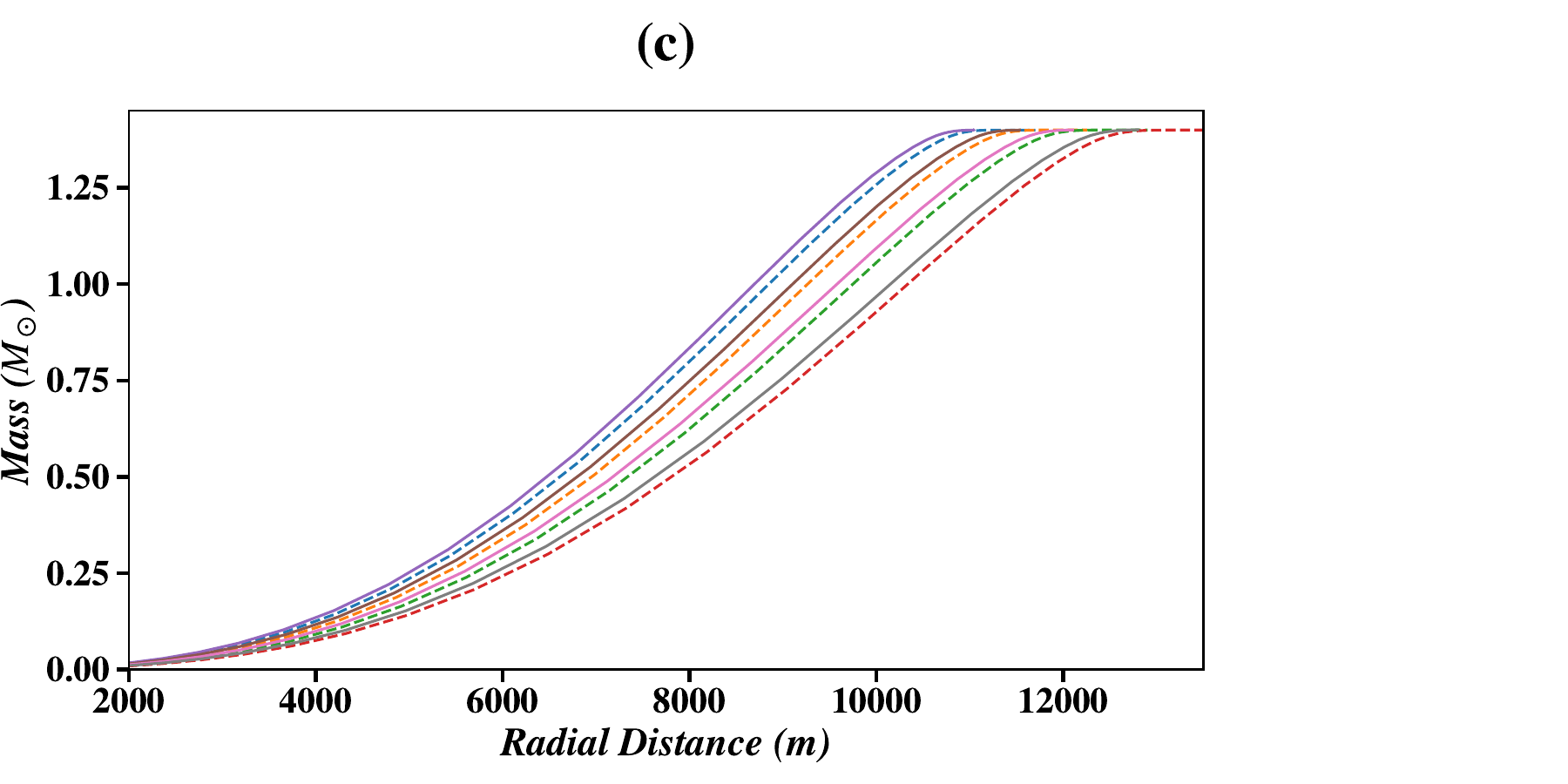}
\caption{ The variation of mass with radial distance \( r \) for different equations of state (EoS): (a) APR, (b) FPS, and (c) SLy, both in the presence (\( B_c = 10^{18} \, \text{Gauss} \)) and absence (\( B_c = 0 \)) of magnetic fields for neutron stars of $1.4 \, M_\odot$. The cases without magnetic fields are labeled as "WoB" (Without Magnetic field), and the cases with magnetic fields are labeled as "WB" (With Magnetic field). The modified gravity parameter \( \lambda \) is set to $0$, \( -\frac{1}{8\pi} \), \( -\frac{2}{8\pi} \), and \( -\frac{3}{8\pi}\).}
\label{fig:masswithr}
\end{figure*}

%%%%%%%%%%%%%%%%%%%%
\begin{figure*}
\includegraphics[scale = 0.29]{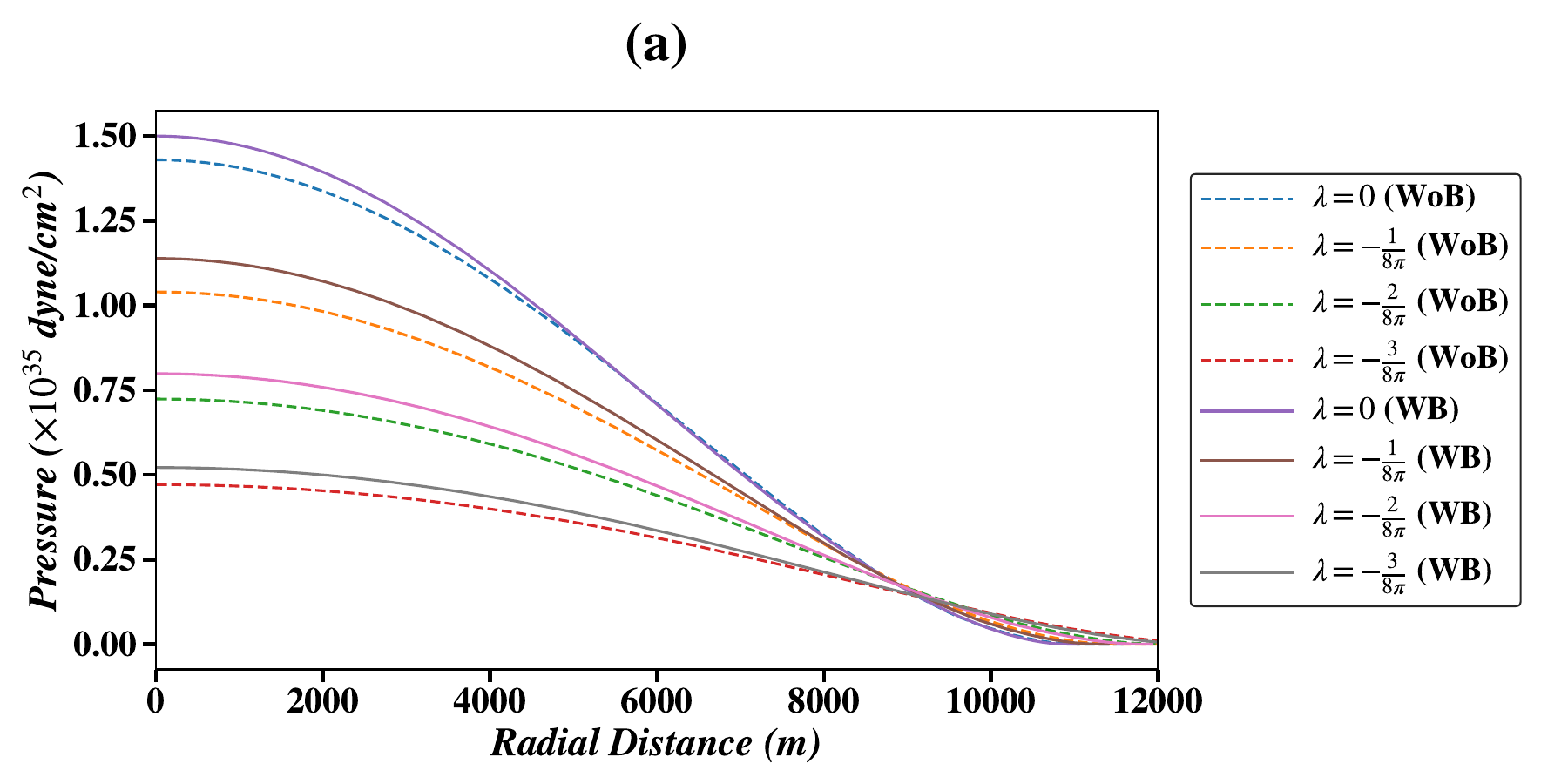}
\includegraphics[scale = 0.29]{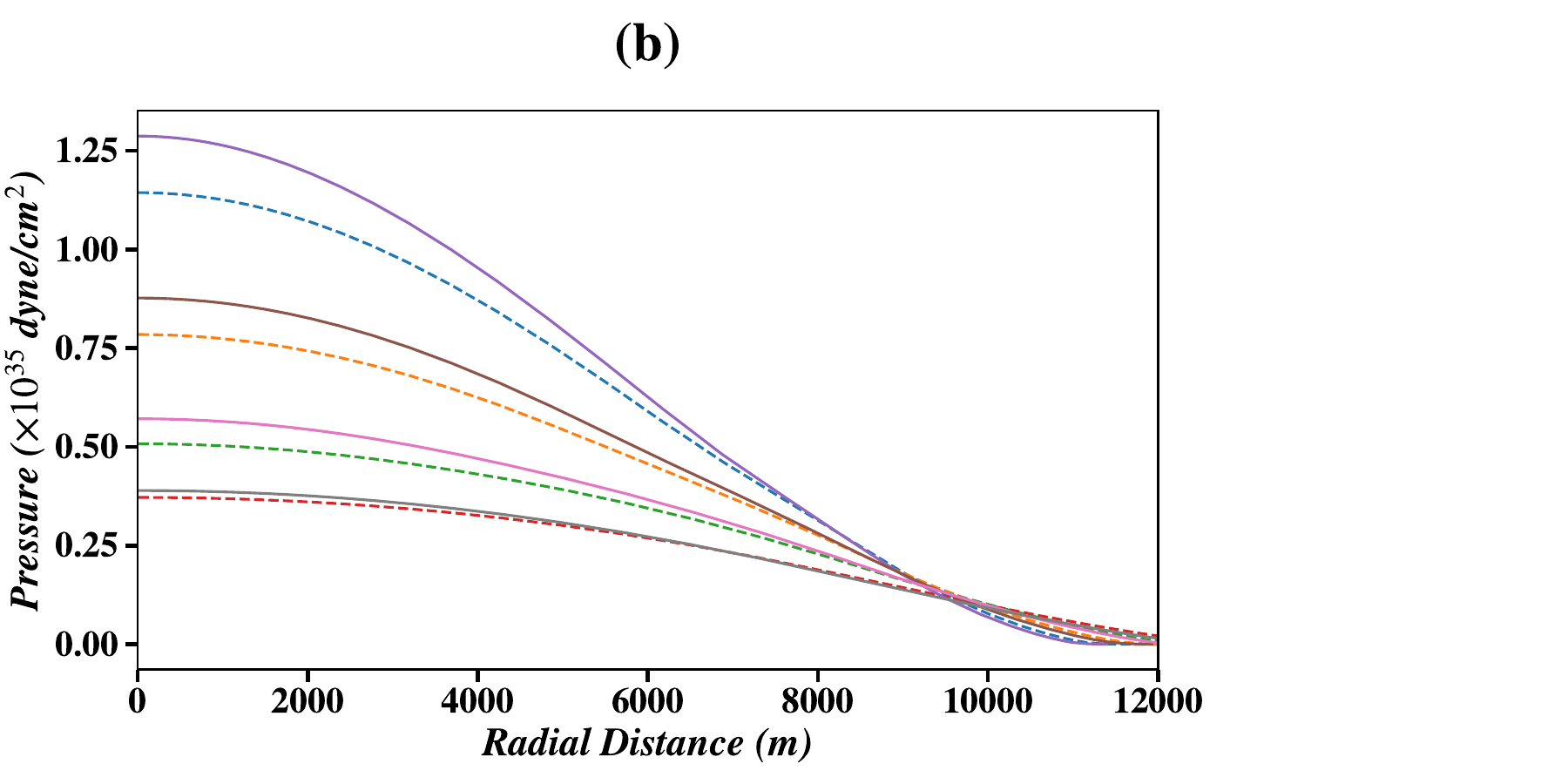}
\includegraphics[scale = 0.29]{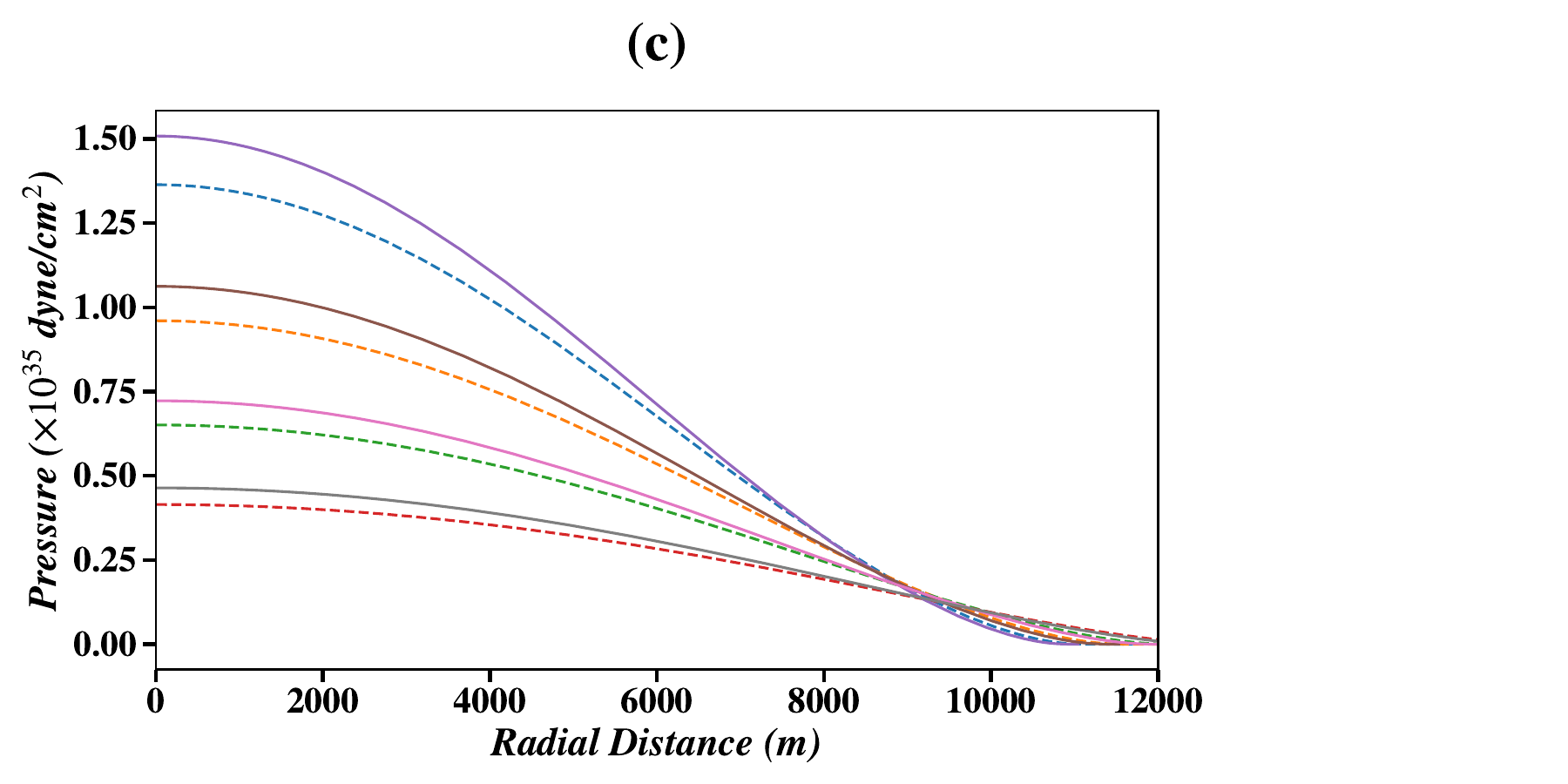}
\caption{The variation of pressure with radial distance \( r \) for different equations of state (EoS): (a) APR, (b) FPS, and (c) SLy, both in the presence (\( B_c = 10^{18} \, \text{Gauss} \)) and absence (\( B_c = 0 \)) of magnetic fields in neutron stars of $1.4 \, M_\odot$. The cases without magnetic fields are labeled as "WoB" (without magnetic field), and the cases with magnetic fields are labeled as "WB" (with magnetic field). The modified gravity parameter \( \lambda \) is set to $0$ ,\(-\frac{1}{8\pi}, -\frac{2}{8\pi}\), and \( -\frac{3}{8\pi}\).}
\label{fig:pressurewithr}
\end{figure*}

\noindent {\it Magnetic Field and EoS Effects on the Mass Profile}  \\

 In Figure (\ref{fig:masswithr}), we show the enclosed mass of the star as a function of the radial distance from the center of a neutron star corresponding to the different values of the $\lambda$ with and without the magnetic field (denoted WB and WoB, respectively, in the figures) for three different values of the modified gravity parameter; $\lambda = 0,-\frac{1}{8\pi},-\frac{2}{8\pi},-\frac{3}{8\pi}$ to see the impact of modified gravity and magnetic field on the mass profile. In Figures~\ref{fig:masswithr}(a), \ref{fig:masswithr}(b) and \ref{fig:masswithr}(c), we have used different EoSs, namely, APR~\cite{akmal1998equation}, FPS~\cite{flowers1976neutrino} and SLy~\cite{broderick2000equation}. During solving the TOV equations, we have considered a central magnetic field of $B_c = 10^{18}$ Gauss to determine the magnetic field at different radial distances inside the NS (magnetic field profile) from the center of the star.\\

\noindent In Figure~\ref{fig:masswithr}(a), we show the variation of the neutron star mass $M/M_{\odot}$ with the radial distance with and without the magnetic field corresponding to the APR EOS.
%%%%%%%%

%%%%%%%%
\noindent As shown in Figure~\ref{fig:masswithr}(a), in the presence of a magnetic field, the mass-radial radius curves shift upward. It suggests that the magnetic field contributes reasonably and requires less radial distance as compared to the non-magnetic field counterpart. This effect is seen consistently on all values of $\lambda$. Consequently, for a typical solar mass of $1.4$, the radius increases when the magnetic field is present compared to the case without it. A similar magnetic field effect is observed for other FPS and SLy EoSs, as shown below.\\

\noindent In our isotropic approximation, the magnetic field contributes to the energy density
$\varepsilon_{B} = \frac{B^{2}}{8\pi}$,
which enters the Tolman–Oppenheimer–Volkoff mass integral as~\cite{shapiro2024black}:
\begin{equation}\label{eq:deltaM}
\Delta M = \frac{1}{c^{2}}\int_{0}^{R}4\pi r^{2}\,\varepsilon_{B}(r)\,\mathrm{d}r\,
\end{equation}
 For a surface dipole strength of \(B_c=10^{18}\)\ Gauss, (implying an average over \(R\simeq12\)km), Eq.~\eqref{eq:deltaM} yields \(\Delta M\approx0.02\,M_{\odot}\)~\cite{cardall2001effects}. 
 Our results shown in Figure~\ref{fig:masswithr} on $\Delta M$  are obtained using the  EoS ( APR, FPS, and SLy) and are in good agreement with the analytic estimate ~\cite{cardall2001effects}. 
 
 % for the APR, FPS and SLy EoSs in Figure:\ref{fig:masswithr} produce \(\Delta M=0.02\)–0.04\(M_{\odot}\) with reasonable agreement with this analytic estimate
 
 This illustrate the modest but clearly visible upward shift of the magnetic (dashed) curves relative to the non-magnetic (solid).
\\

%%%%%%%%%%%%%%%%%%%
\noindent{\it Modified Gravity and EOS Effects on the Mass Profile}\\

Further, the influence of  $\lambda$ is observed under both the magnetic field and choosen EoSs. The neutron star appears to be more compact for $\lambda = -1/8\pi$. However, as $\lambda$ changes towards more negative values, the radius for a given mass consistently increases, indicating that weaker effective gravity results in a "fluffier" neutron star. This trend is evident in both cases, in the presence and absence of the magnetic field. This effect is seen to be more pronounced in the $B = 0$ case, where gravitational binding is reduced and no additional magnetic pressure is available to stabilize the neutron star.\\

\noindent {\it Magnetic Field and EoS Effects on the Pressure Profile}\\

Figure \ref{fig:pressurewithr} shows the influence of modified gravity and magnetic field on the pressure versus radial distance curves, as shown in Figures \ref{fig:pressurewithr}(a), \ref{fig:pressurewithr}(b), and \ref{fig:pressurewithr}(c) for APR, FPS, and SLy EoSs, respectively.\\ 
It is clear from Figure \ref{fig:pressurewithr}(a) that the magnetic field significantly impacts the pressure at lower values of radial distances from the center of the star. This effect of the magnetic field decreases as we move toward the surface of the star. \\
It is also evident from the figure that the magnetic field impact becomes more dominant for lower negative values of the $\lambda$. Upon comparing Figures \ref{fig:pressurewithr}(a), \ref{fig:pressurewithr}(b), and \ref{fig:pressurewithr}(c), we found that the qualitative variation of pressure versus radial distance does not change significantly when we move from APR to FPS or FPS to SLy EoS. \\ 

\noindent {\it Modified Gravity and EOS Effects on Pressure Profile}\\

% As $\lambda$ shifts towards more negative values from $-\frac{1}{8\pi},-\frac{2}{8\pi},-\frac{3}{8\pi}$, the pressure decreases and its gradient with respect to the radial distance becomes less steeper, implying that the reduction in gravitational pull leads to lower equilibrium pressures within the neutron star. This suggests that stars with more negative values $\lambda$ are less gravitationally bound, making it challenging to sustain high internal pressures, which aligns with the increased radial distance observed in the mass-radial distance plots. \\

As \(\lambda\) becomes more negative (from \(-\tfrac{1}{8\pi}\) to \(-\tfrac{3}{8\pi}\)), the pressure profile decreases and its radial gradient becomes less steep.

This behavior reflects a weaker effective gravitational pull in the equilibrium equations.
Consequently, the star requires a lower internal pressure to balance gravity at a given radius and therefore configurations are less tightly bound.
Continually, the mass-radius sequences shift to larger radii as \(\lambda\) becomes more negative.\\

%%%%%%%%%%%%%%%%%%%%%%%%%%%%%%%%%%%%%   

\section{Conclusions and Future Outlook}
\label{conc}
\noindent 
In this study, we have investigated the structural properties of neutron stars within the linear modified gravity framework \( f(R, T) = R + 2\lambda \kappa T \), incorporating the combined effects of strong magnetic fields and realistic nuclear equations of state (APR, FPS, and SLy). The modified TOV equations are solved numerically for both non-magnetized (WoB) and magnetized (WB) stars,  considering a central magnetic field of \( B_{c} = 10^{18} \, \text{Gauss} \). Through this self-consistent framework, the study explored how matter–geometry coupling and magnetic field strength jointly regulate stellar equilibrium and internal structure. A few key findings of our present work are listed as follows;

\begin{itemize}
\item
A negative coupling parameter \( \lambda \) systematically, shifting the mass–radius (\( M-R \)) relationship upward and rightward relative to the predictions of General Relativity (GR). For \( \lambda = -3/(8\pi) \), the APR sequence reaches a maximum mass of approximately \( 2.7\, M_{\odot} \), while SLy and FPS exhibit slightly smaller but consistent trends. It indicates that the effective gravitational coupling weakens for negative \( \lambda \), allowing equilibrium configurations with larger radii and lower central densities.

\item
Introducing a strong internal magnetic field (\( B_{c} = 10^{18}\,\text{Gauss} \)) slightly increases the net gravitational mass (\( \Delta M \approx 0.02\, M_{\odot} \)) through magnetic energy contributions but produces only minor modifications in radius and pressure. The field remains insufficient to break spherical symmetry, supporting the validity of the isotropic pressure approximation in this regime.

\item
The predicted \( M-R \) curves intersect the mass bands of massive pulsars such as PSR~J0348+0432 and PSR~J0740+6620, align with the NICER posteriors for J0030+0451, and remain within the GW170817 tidal deformability window (\( R_{1.4} \sim 11-13\,\text{km} \)). These results confirm that even with strong magnetic fields and moderate values of 
\( \lambda \), the model remains compatible with multimessenger constraints and current astrophysical data.

\item 
The analysis of mass and pressure distributions shows that magnetic fields primarily influence the central core, whereas modified gravity alters the global stellar profile. Their combined effect yields modest deviations from GR predictions, particularly near the maximum-mass configuration, while preserving overall stability within the turning-point criterion.
\end{itemize}

Building upon the present outcomes, a few promising directions for further investigation are identified:

\begin{itemize}
\item
Future studies may extend this framework, to compute tidal Love numbers \( k_2(M) \) and deformabilities \( \Lambda(M) \) for both magnetized and non-magnetized configurations. Such results will enable direct confrontation with gravitational-wave observations from binary neutron star mergers (e.g., GW170817) and allow tests of potential modifications in the empirical I–Love–Q relations under \( f(R, T) \) gravity.

\item
While pressure isotropy holds reasonably well up to \( B_{c} \sim 10^{18}\,\text{Gauss} \), more realistic models require the inclusion of anisotropy induced by mixed poloidal–toroidal magnetic field geometries and rotation. Incorporating these effects will enable the study of shape deformations, stability boundaries, and magneto-rotational couplings in compact stars governed by modified gravity.

\item
Introducing higher-order curvature terms such as \( f(R, T) = R + \alpha R^2 + 2\lambda\kappa T \) could reveal additional departures from GR in ultra-dense regimes. Examining such hybrid extensions will clarify whether curvature-driven or matter–coupled modifications dominate neutron star observables.

\item
Coupling the present framework with temperature-dependent equations of state and neutrino-cooling dynamics will allow the study of thermal evolution, magnetic decay, and the long-term stability of magnetars. These investigations may provide observable signatures connecting modified gravity with the thermal and magnetic lifecycles of compact stars.
\end{itemize}

\bigskip
\noindent
In summary, the present study demonstrates that moderate matter–geometry coupling within 
\( f(R, T) \) gravity can stiffen the neutron star equation of state while maintaining consistency with all known observational constraints. Although magnetic fields contribute only secondary corrections, their interplay with modified gravity provides a meaningful avenue to probe deviations from GR in the strong-field regime. Future high-precision gravitational-wave and X-ray observations will be crucial for constraining the coupling parameter \( \lambda \) and testing the viability of modified gravity in ultra-dense astrophysical environments.

\section {Acknowledgments}
One of the authors, Charul Rathod (CR), is grateful to DST-New Delhi for providing the financial support as a DST-INSPIRE fellowship. CR also acknowledges Birla Institute of Technology and Science - Pilani, Pilani Campus, Pilani - 333031 (Rajasthan) for providing the research facilities and the required administrative support. We would like to thank Dr. Captain Rituraj Singh for clarifying the doubts and his valuable suggestions.  

%%%%%%%%%%%%%%%%%%%%%%%%%%%%%%%%%
\bibliographystyle{apsrev4-2} 
\bibliography{references}

%apsrev4-2.bst 2019-01-14 (MD) hand-edited version of apsrev4-1.bst
%Control: key (0)
%Control: author (72) initials jnrlst
%Control: editor formatted (1) identically to author
%Control: production of article title (-1) disabled
%Control: page (0) single
%Control: year (1) truncated
%Control: production of eprint (0) enabled
\begin{thebibliography}{47}%
\makeatletter
\providecommand \@ifxundefined [1]{%
 \@ifx{#1\undefined}
}%
\providecommand \@ifnum [1]{%
 \ifnum #1\expandafter \@firstoftwo
 \else \expandafter \@secondoftwo
 \fi
}%
\providecommand \@ifx [1]{%
 \ifx #1\expandafter \@firstoftwo
 \else \expandafter \@secondoftwo
 \fi
}%
\providecommand \natexlab [1]{#1}%
\providecommand \enquote  [1]{``#1''}%
\providecommand \bibnamefont  [1]{#1}%
\providecommand \bibfnamefont [1]{#1}%
\providecommand \citenamefont [1]{#1}%
\providecommand \href@noop [0]{\@secondoftwo}%
\providecommand \href [0]{\begingroup \@sanitize@url \@href}%
\providecommand \@href[1]{\@@startlink{#1}\@@href}%
\providecommand \@@href[1]{\endgroup#1\@@endlink}%
\providecommand \@sanitize@url [0]{\catcode `\\12\catcode `\$12\catcode
  `\&12\catcode `\#12\catcode `\^12\catcode `\_12\catcode `\%12\relax}%
\providecommand \@@startlink[1]{}%
\providecommand \@@endlink[0]{}%
\providecommand \url  [0]{\begingroup\@sanitize@url \@url }%
\providecommand \@url [1]{\endgroup\@href {#1}{\urlprefix }}%
\providecommand \urlprefix  [0]{URL }%
\providecommand \Eprint [0]{\href }%
\providecommand \doibase [0]{https://doi.org/}%
\providecommand \selectlanguage [0]{\@gobble}%
\providecommand \bibinfo  [0]{\@secondoftwo}%
\providecommand \bibfield  [0]{\@secondoftwo}%
\providecommand \translation [1]{[#1]}%
\providecommand \BibitemOpen [0]{}%
\providecommand \bibitemStop [0]{}%
\providecommand \bibitemNoStop [0]{.\EOS\space}%
\providecommand \EOS [0]{\spacefactor3000\relax}%
\providecommand \BibitemShut  [1]{\csname bibitem#1\endcsname}%
\let\auto@bib@innerbib\@empty
%</preamble>
\bibitem [{\citenamefont {Lattimer}\ and\ \citenamefont
  {Prakash}(2004)}]{lattimer2004physics}%
  \BibitemOpen
  \bibfield  {author} {\bibinfo {author} {\bibfnamefont {J.~M.}\ \bibnamefont
  {Lattimer}}\ and\ \bibinfo {author} {\bibfnamefont {M.}~\bibnamefont
  {Prakash}},\ }\href@noop {} {\bibfield  {journal} {\bibinfo  {journal}
  {Science}\ }\textbf {\bibinfo {volume} {304}},\ \bibinfo {pages} {536}
  (\bibinfo {year} {2004})}\BibitemShut {NoStop}%
\bibitem [{\citenamefont {Dexheimer}\ \emph {et~al.}(2017)\citenamefont
  {Dexheimer}, \citenamefont {Franzon}, \citenamefont {Gomes}, \citenamefont
  {Farias},\ and\ \citenamefont {Avancini}}]{dexheimer2017magnetic}%
  \BibitemOpen
  \bibfield  {author} {\bibinfo {author} {\bibfnamefont {V.}~\bibnamefont
  {Dexheimer}}, \bibinfo {author} {\bibfnamefont {B.}~\bibnamefont {Franzon}},
  \bibinfo {author} {\bibfnamefont {R.}~\bibnamefont {Gomes}}, \bibinfo
  {author} {\bibfnamefont {R.}~\bibnamefont {Farias}},\ and\ \bibinfo {author}
  {\bibfnamefont {S.}~\bibnamefont {Avancini}},\ }\href@noop {} {\bibfield
  {journal} {\bibinfo  {journal} {Astronomische Nachrichten}\ }\textbf
  {\bibinfo {volume} {338}},\ \bibinfo {pages} {1052} (\bibinfo {year}
  {2017})}\BibitemShut {NoStop}%
\bibitem [{\citenamefont {N{\"a}ttil{\"a}}\ and\ \citenamefont
  {Kajava}(2022)}]{nattila2022fundamental}%
  \BibitemOpen
  \bibfield  {author} {\bibinfo {author} {\bibfnamefont {J.}~\bibnamefont
  {N{\"a}ttil{\"a}}}\ and\ \bibinfo {author} {\bibfnamefont {J.~J.}\
  \bibnamefont {Kajava}}\ }(\bibinfo  {publisher} {Springer},\ \bibinfo {year}
  {2022})\ pp.\ \bibinfo {pages} {1--53}\BibitemShut {NoStop}%
\bibitem [{\citenamefont {Hehl}(1997)}]{hehl1997alternative}%
  \BibitemOpen
  \bibfield  {author} {\bibinfo {author} {\bibfnamefont {F.~W.}\ \bibnamefont
  {Hehl}},\ }\href@noop {} {\bibfield  {journal} {\bibinfo  {journal} {arXiv
  preprint gr-qc/9712096}\ } (\bibinfo {year} {1997})}\BibitemShut {NoStop}%
\bibitem [{\citenamefont {Kunz}(2023)}]{kunz2023neutron}%
  \BibitemOpen
  \bibfield  {author} {\bibinfo {author} {\bibfnamefont {J.}~\bibnamefont
  {Kunz}}\ }(\bibinfo  {publisher} {Springer},\ \bibinfo {year} {2023})\ pp.\
  \bibinfo {pages} {293--313}\BibitemShut {NoStop}%
\bibitem [{\citenamefont {Einstein}(1915)}]{einstein1915field}%
  \BibitemOpen
  \bibfield  {author} {\bibinfo {author} {\bibfnamefont {A.}~\bibnamefont
  {Einstein}},\ }\href@noop {} {\bibfield  {journal} {\bibinfo  {journal}
  {Sitzungsber. Preuss. Akad. Wiss. Berlin (Math. Phys.)}\ }\textbf {\bibinfo
  {volume} {1915}},\ \bibinfo {pages} {844} (\bibinfo {year}
  {1915})}\BibitemShut {NoStop}%
\bibitem [{\citenamefont {Sotiriou}\ and\ \citenamefont
  {Faraoni}(2010)}]{sotiriou2010f}%
  \BibitemOpen
  \bibfield  {author} {\bibinfo {author} {\bibfnamefont {T.~P.}\ \bibnamefont
  {Sotiriou}}\ and\ \bibinfo {author} {\bibfnamefont {V.}~\bibnamefont
  {Faraoni}},\ }\href@noop {} {\bibfield  {journal} {\bibinfo  {journal}
  {Reviews of Modern Physics}\ }\textbf {\bibinfo {volume} {82}},\ \bibinfo
  {pages} {451} (\bibinfo {year} {2010})}\BibitemShut {NoStop}%
\bibitem [{\citenamefont {De~Felice}\ and\ \citenamefont
  {Tsujikawa}(2010)}]{de2010f}%
  \BibitemOpen
  \bibfield  {author} {\bibinfo {author} {\bibfnamefont {A.}~\bibnamefont
  {De~Felice}}\ and\ \bibinfo {author} {\bibfnamefont {S.}~\bibnamefont
  {Tsujikawa}},\ }\href@noop {} {\bibfield  {journal} {\bibinfo  {journal}
  {Living Reviews in Relativity}\ }\textbf {\bibinfo {volume} {13}},\ \bibinfo
  {pages} {1} (\bibinfo {year} {2010})}\BibitemShut {NoStop}%
\bibitem [{\citenamefont {Bauer}\ \emph {et~al.}(2022)\citenamefont {Bauer},
  \citenamefont {C{\'a}rdenas-Avenda{\~n}o}, \citenamefont {Gammie},\ and\
  \citenamefont {Yunes}}]{bauer2022spherical}%
  \BibitemOpen
  \bibfield  {author} {\bibinfo {author} {\bibfnamefont {A.~M.}\ \bibnamefont
  {Bauer}}, \bibinfo {author} {\bibfnamefont {A.}~\bibnamefont
  {C{\'a}rdenas-Avenda{\~n}o}}, \bibinfo {author} {\bibfnamefont {C.~F.}\
  \bibnamefont {Gammie}},\ and\ \bibinfo {author} {\bibfnamefont
  {N.}~\bibnamefont {Yunes}},\ }\href@noop {} {\bibfield  {journal} {\bibinfo
  {journal} {The Astrophysical Journal}\ }\textbf {\bibinfo {volume} {925}},\
  \bibinfo {pages} {119} (\bibinfo {year} {2022})}\BibitemShut {NoStop}%
\bibitem [{\citenamefont {Alex}\ and\ \citenamefont
  {Reinhart}(2020)}]{alex2020covariant}%
  \BibitemOpen
  \bibfield  {author} {\bibinfo {author} {\bibfnamefont {N.}~\bibnamefont
  {Alex}}\ and\ \bibinfo {author} {\bibfnamefont {T.}~\bibnamefont
  {Reinhart}},\ }\href@noop {} {\bibfield  {journal} {\bibinfo  {journal}
  {Physical Review D}\ }\textbf {\bibinfo {volume} {101}},\ \bibinfo {pages}
  {084025} (\bibinfo {year} {2020})}\BibitemShut {NoStop}%
\bibitem [{\citenamefont {Nobleson}\ \emph {et~al.}(2022)\citenamefont
  {Nobleson}, \citenamefont {Ali},\ and\ \citenamefont
  {Banik}}]{nobleson2022comparison}%
  \BibitemOpen
  \bibfield  {author} {\bibinfo {author} {\bibfnamefont {K.}~\bibnamefont
  {Nobleson}}, \bibinfo {author} {\bibfnamefont {A.}~\bibnamefont {Ali}},\ and\
  \bibinfo {author} {\bibfnamefont {S.}~\bibnamefont {Banik}},\ }\href@noop {}
  {\bibfield  {journal} {\bibinfo  {journal} {The European Physical Journal C}\
  }\textbf {\bibinfo {volume} {82}},\ \bibinfo {pages} {32} (\bibinfo {year}
  {2022})}\BibitemShut {NoStop}%
\bibitem [{\citenamefont {Harko}\ \emph {et~al.}(2011)\citenamefont {Harko},
  \citenamefont {Lobo}, \citenamefont {Nojiri},\ and\ \citenamefont
  {Odintsov}}]{harko2011f}%
  \BibitemOpen
  \bibfield  {author} {\bibinfo {author} {\bibfnamefont {T.}~\bibnamefont
  {Harko}}, \bibinfo {author} {\bibfnamefont {F.~S.}\ \bibnamefont {Lobo}},
  \bibinfo {author} {\bibfnamefont {S.}~\bibnamefont {Nojiri}},\ and\ \bibinfo
  {author} {\bibfnamefont {S.~D.}\ \bibnamefont {Odintsov}},\ }\href@noop {}
  {\bibfield  {journal} {\bibinfo  {journal} {Physical Review D—Particles,
  Fields, Gravitation, and Cosmology}\ }\textbf {\bibinfo {volume} {84}},\
  \bibinfo {pages} {024020} (\bibinfo {year} {2011})}\BibitemShut {NoStop}%
\bibitem [{\citenamefont {Mahapatra}\ and\ \citenamefont
  {Das}(2024)}]{mahapatra2024neutron}%
  \BibitemOpen
  \bibfield  {author} {\bibinfo {author} {\bibfnamefont {P.}~\bibnamefont
  {Mahapatra}}\ and\ \bibinfo {author} {\bibfnamefont {P.~K.}\ \bibnamefont
  {Das}},\ }\href@noop {} {\bibfield  {journal} {\bibinfo  {journal} {arXiv
  preprint}\ } (\bibinfo {year} {2024})},\ \Eprint
  {https://arxiv.org/abs/2401.01321} {arXiv:2401.01321} \BibitemShut {NoStop}%
\bibitem [{\citenamefont {Carvalho}\ \emph {et~al.}(2017)\citenamefont
  {Carvalho}, \citenamefont {Lobato}, \citenamefont {Moraes}, \citenamefont
  {Arba{\~n}il}, \citenamefont {Otoniel}, \citenamefont {Marinho},\ and\
  \citenamefont {Malheiro}}]{carvalho2017stellar}%
  \BibitemOpen
  \bibfield  {author} {\bibinfo {author} {\bibfnamefont {G.}~\bibnamefont
  {Carvalho}}, \bibinfo {author} {\bibfnamefont {R.}~\bibnamefont {Lobato}},
  \bibinfo {author} {\bibfnamefont {P.}~\bibnamefont {Moraes}}, \bibinfo
  {author} {\bibfnamefont {J.~D.}\ \bibnamefont {Arba{\~n}il}}, \bibinfo
  {author} {\bibfnamefont {E.}~\bibnamefont {Otoniel}}, \bibinfo {author}
  {\bibfnamefont {R.}~\bibnamefont {Marinho}},\ and\ \bibinfo {author}
  {\bibfnamefont {M.}~\bibnamefont {Malheiro}},\ }\href@noop {} {\bibfield
  {journal} {\bibinfo  {journal} {The European Physical Journal C}\ }\textbf
  {\bibinfo {volume} {77}},\ \bibinfo {pages} {1} (\bibinfo {year}
  {2017})}\BibitemShut {NoStop}%
\bibitem [{\citenamefont {Lobato}\ \emph {et~al.}(2020)\citenamefont {Lobato},
  \citenamefont {Louren{\c{c}}o}, \citenamefont {Moraes}, \citenamefont
  {Lenzi}, \citenamefont {De~Avellar}, \citenamefont {De~Paula}, \citenamefont
  {Dutra},\ and\ \citenamefont {Malheiro}}]{lobato2020neutron}%
  \BibitemOpen
  \bibfield  {author} {\bibinfo {author} {\bibfnamefont {R.}~\bibnamefont
  {Lobato}}, \bibinfo {author} {\bibfnamefont {O.}~\bibnamefont
  {Louren{\c{c}}o}}, \bibinfo {author} {\bibfnamefont {P.}~\bibnamefont
  {Moraes}}, \bibinfo {author} {\bibfnamefont {C.}~\bibnamefont {Lenzi}},
  \bibinfo {author} {\bibfnamefont {M.}~\bibnamefont {De~Avellar}}, \bibinfo
  {author} {\bibfnamefont {W.}~\bibnamefont {De~Paula}}, \bibinfo {author}
  {\bibfnamefont {M.}~\bibnamefont {Dutra}},\ and\ \bibinfo {author}
  {\bibfnamefont {M.}~\bibnamefont {Malheiro}},\ }\href@noop {} {\bibfield
  {journal} {\bibinfo  {journal} {Journal of Cosmology and Astroparticle
  Physics}\ }\textbf {\bibinfo {volume} {2020}}\bibinfo  {number} { (12)},\
  \bibinfo {pages} {039}}\BibitemShut {NoStop}%
\bibitem [{\citenamefont {Ferrer}\ \emph {et~al.}(2010)\citenamefont {Ferrer},
  \citenamefont {de~La~Incera}, \citenamefont {Keith}, \citenamefont
  {Portillo},\ and\ \citenamefont {Springsteen}}]{ferrer2010equation}%
  \BibitemOpen
\bibfield  {number} {  }\bibfield  {author} {\bibinfo {author} {\bibfnamefont
  {E.~J.}\ \bibnamefont {Ferrer}}, \bibinfo {author} {\bibfnamefont
  {V.}~\bibnamefont {de~La~Incera}}, \bibinfo {author} {\bibfnamefont {J.~P.}\
  \bibnamefont {Keith}}, \bibinfo {author} {\bibfnamefont {I.}~\bibnamefont
  {Portillo}},\ and\ \bibinfo {author} {\bibfnamefont {P.~L.}\ \bibnamefont
  {Springsteen}},\ }\href@noop {} {\bibfield  {journal} {\bibinfo  {journal}
  {Physical Review C—Nuclear Physics}\ }\textbf {\bibinfo {volume} {82}},\
  \bibinfo {pages} {065802} (\bibinfo {year} {2010})}\BibitemShut {NoStop}%
\bibitem [{\citenamefont {Ferrer}\ and\ \citenamefont
  {Hackebill}(2022)}]{ferrer2022hadron}%
  \BibitemOpen
  \bibfield  {author} {\bibinfo {author} {\bibfnamefont {E.~J.}\ \bibnamefont
  {Ferrer}}\ and\ \bibinfo {author} {\bibfnamefont {A.}~\bibnamefont
  {Hackebill}},\ }\href@noop {} {\bibfield  {journal} {\bibinfo  {journal}
  {International journal of modern physics A}\ }\textbf {\bibinfo {volume}
  {37}},\ \bibinfo {pages} {2250048} (\bibinfo {year} {2022})}\BibitemShut
  {NoStop}%
\bibitem [{\citenamefont {Chatterjee}\ \emph {et~al.}(2019)\citenamefont
  {Chatterjee}, \citenamefont {Novak},\ and\ \citenamefont
  {Oertel}}]{chatterjee2019magnetic}%
  \BibitemOpen
  \bibfield  {author} {\bibinfo {author} {\bibfnamefont {D.}~\bibnamefont
  {Chatterjee}}, \bibinfo {author} {\bibfnamefont {J.}~\bibnamefont {Novak}},\
  and\ \bibinfo {author} {\bibfnamefont {M.}~\bibnamefont {Oertel}},\
  }\href@noop {} {\bibfield  {journal} {\bibinfo  {journal} {Physical Review
  C}\ }\textbf {\bibinfo {volume} {99}},\ \bibinfo {pages} {055811} (\bibinfo
  {year} {2019})}\BibitemShut {NoStop}%
\bibitem [{\citenamefont {Kaspi}\ and\ \citenamefont
  {Beloborodov}(2017)}]{kaspi2017magnetars}%
  \BibitemOpen
  \bibfield  {author} {\bibinfo {author} {\bibfnamefont {V.~M.}\ \bibnamefont
  {Kaspi}}\ and\ \bibinfo {author} {\bibfnamefont {A.~M.}\ \bibnamefont
  {Beloborodov}},\ }\href@noop {} {\bibfield  {journal} {\bibinfo  {journal}
  {Annual Review of Astronomy and Astrophysics}\ }\textbf {\bibinfo {volume}
  {55}},\ \bibinfo {pages} {261} (\bibinfo {year} {2017})}\BibitemShut
  {NoStop}%
\bibitem [{\citenamefont {Yadav}\ \emph {et~al.}(2024)\citenamefont {Yadav},
  \citenamefont {Mishra}, \citenamefont {Sarkar},\ and\ \citenamefont
  {Singh}}]{yadav2024thermal}%
  \BibitemOpen
  \bibfield  {author} {\bibinfo {author} {\bibfnamefont {S.}~\bibnamefont
  {Yadav}}, \bibinfo {author} {\bibfnamefont {M.}~\bibnamefont {Mishra}},
  \bibinfo {author} {\bibfnamefont {T.~G.}\ \bibnamefont {Sarkar}},\ and\
  \bibinfo {author} {\bibfnamefont {C.~R.}\ \bibnamefont {Singh}},\ }\href@noop
  {} {\bibfield  {journal} {\bibinfo  {journal} {The European Physical Journal
  C}\ }\textbf {\bibinfo {volume} {84}},\ \bibinfo {pages} {225} (\bibinfo
  {year} {2024})}\BibitemShut {NoStop}%
\bibitem [{\citenamefont {Akmal}\ \emph {et~al.}(1998)\citenamefont {Akmal},
  \citenamefont {Pandharipande},\ and\ \citenamefont
  {Ravenhall}}]{akmal1998equation}%
  \BibitemOpen
  \bibfield  {author} {\bibinfo {author} {\bibfnamefont {A.}~\bibnamefont
  {Akmal}}, \bibinfo {author} {\bibfnamefont {V.}~\bibnamefont
  {Pandharipande}},\ and\ \bibinfo {author} {\bibfnamefont {D.}~\bibnamefont
  {Ravenhall}},\ }\href@noop {} {\bibfield  {journal} {\bibinfo  {journal}
  {Physical Review C}\ }\textbf {\bibinfo {volume} {58}},\ \bibinfo {pages}
  {1804} (\bibinfo {year} {1998})}\BibitemShut {NoStop}%
\bibitem [{\citenamefont {Schneider}\ \emph {et~al.}(2019)\citenamefont
  {Schneider}, \citenamefont {Constantinou}, \citenamefont {Muccioli},\ and\
  \citenamefont {Prakash}}]{schneider2019akmal}%
  \BibitemOpen
  \bibfield  {author} {\bibinfo {author} {\bibfnamefont {A.~S.}\ \bibnamefont
  {Schneider}}, \bibinfo {author} {\bibfnamefont {C.}~\bibnamefont
  {Constantinou}}, \bibinfo {author} {\bibfnamefont {B.}~\bibnamefont
  {Muccioli}},\ and\ \bibinfo {author} {\bibfnamefont {M.}~\bibnamefont
  {Prakash}},\ }\href@noop {} {\bibfield  {journal} {\bibinfo  {journal}
  {Physical Review C}\ }\textbf {\bibinfo {volume} {100}},\ \bibinfo {pages}
  {025803} (\bibinfo {year} {2019})}\BibitemShut {NoStop}%
\bibitem [{\citenamefont {Flowers}\ \emph {et~al.}(1976)\citenamefont
  {Flowers}, \citenamefont {Ruderman},\ and\ \citenamefont
  {Sutherland}}]{flowers1976neutrino}%
  \BibitemOpen
  \bibfield  {author} {\bibinfo {author} {\bibfnamefont {E.}~\bibnamefont
  {Flowers}}, \bibinfo {author} {\bibfnamefont {M.}~\bibnamefont {Ruderman}},\
  and\ \bibinfo {author} {\bibfnamefont {P.}~\bibnamefont {Sutherland}},\
  }\href@noop {} {\bibfield  {journal} {\bibinfo  {journal} {Astrophysical
  Journal, vol. 205, Apr. 15, 1976, pt. 1, p. 541-544.}\ }\textbf {\bibinfo
  {volume} {205}},\ \bibinfo {pages} {541} (\bibinfo {year}
  {1976})}\BibitemShut {NoStop}%
\bibitem [{\citenamefont {Tolman}(1939)}]{tolman1939static}%
  \BibitemOpen
  \bibfield  {author} {\bibinfo {author} {\bibfnamefont {R.~C.}\ \bibnamefont
  {Tolman}},\ }\href@noop {} {\bibfield  {journal} {\bibinfo  {journal}
  {Physical Review}\ }\textbf {\bibinfo {volume} {55}},\ \bibinfo {pages} {364}
  (\bibinfo {year} {1939})}\BibitemShut {NoStop}%
\bibitem [{\citenamefont {Oppenheimer}\ and\ \citenamefont
  {Volkoff}(1939)}]{oppenheimer1939massive}%
  \BibitemOpen
  \bibfield  {author} {\bibinfo {author} {\bibfnamefont {J.~R.}\ \bibnamefont
  {Oppenheimer}}\ and\ \bibinfo {author} {\bibfnamefont {G.~M.}\ \bibnamefont
  {Volkoff}},\ }\href@noop {} {\bibfield  {journal} {\bibinfo  {journal}
  {Physical Review}\ }\textbf {\bibinfo {volume} {55}},\ \bibinfo {pages} {374}
  (\bibinfo {year} {1939})}\BibitemShut {NoStop}%
\bibitem [{\citenamefont {Hartle}\ and\ \citenamefont
  {Thorne}(1968)}]{hartle1968slowly}%
  \BibitemOpen
  \bibfield  {author} {\bibinfo {author} {\bibfnamefont {J.~B.}\ \bibnamefont
  {Hartle}}\ and\ \bibinfo {author} {\bibfnamefont {K.~S.}\ \bibnamefont
  {Thorne}},\ }\href@noop {} {\bibfield  {journal} {\bibinfo  {journal}
  {Astrophysical Journal, vol. 153, p. 807}\ }\textbf {\bibinfo {volume}
  {153}},\ \bibinfo {pages} {807} (\bibinfo {year} {1968})}\BibitemShut
  {NoStop}%
\bibitem [{\citenamefont {Schleich}\ and\ \citenamefont
  {Witt}(2009)}]{schleich2009does}%
  \BibitemOpen
  \bibfield  {author} {\bibinfo {author} {\bibfnamefont {K.}~\bibnamefont
  {Schleich}}\ and\ \bibinfo {author} {\bibfnamefont {D.~M.}\ \bibnamefont
  {Witt}},\ }\href@noop {} {\bibfield  {journal} {\bibinfo  {journal} {arXiv
  preprint arXiv:0910.5194}\ } (\bibinfo {year} {2009})}\BibitemShut {NoStop}%
\bibitem [{\citenamefont {Broderick}\ \emph {et~al.}(2000)\citenamefont
  {Broderick}, \citenamefont {Prakash},\ and\ \citenamefont
  {Lattimer}}]{broderick2000equation}%
  \BibitemOpen
  \bibfield  {author} {\bibinfo {author} {\bibfnamefont {A.}~\bibnamefont
  {Broderick}}, \bibinfo {author} {\bibfnamefont {M.}~\bibnamefont {Prakash}},\
  and\ \bibinfo {author} {\bibfnamefont {J.}~\bibnamefont {Lattimer}},\
  }\href@noop {} {\bibfield  {journal} {\bibinfo  {journal} {The Astrophysical
  Journal}\ }\textbf {\bibinfo {volume} {537}},\ \bibinfo {pages} {351}
  (\bibinfo {year} {2000})}\BibitemShut {NoStop}%
\bibitem [{\citenamefont {Pretel}\ \emph {et~al.}(2021)\citenamefont {Pretel},
  \citenamefont {Jor{\'a}s}, \citenamefont {Reis},\ and\ \citenamefont
  {Arba{\~n}il}}]{pretel2021neutron}%
  \BibitemOpen
  \bibfield  {author} {\bibinfo {author} {\bibfnamefont {J.~M.}\ \bibnamefont
  {Pretel}}, \bibinfo {author} {\bibfnamefont {S.~E.}\ \bibnamefont
  {Jor{\'a}s}}, \bibinfo {author} {\bibfnamefont {R.~R.}\ \bibnamefont
  {Reis}},\ and\ \bibinfo {author} {\bibfnamefont {J.~D.}\ \bibnamefont
  {Arba{\~n}il}},\ }\href@noop {} {\bibfield  {journal} {\bibinfo  {journal}
  {Journal of Cosmology and Astroparticle Physics}\ }\textbf {\bibinfo {volume}
  {2021}}\bibinfo  {number} { (08)},\ \bibinfo {pages} {055}}\BibitemShut
  {NoStop}%
\bibitem [{\citenamefont {Moraes}\ \emph {et~al.}(2016)\citenamefont {Moraes},
  \citenamefont {Arba{\~n}il},\ and\ \citenamefont
  {Malheiro}}]{moraes2016stellar}%
  \BibitemOpen
\bibfield  {number} {  }\bibfield  {author} {\bibinfo {author} {\bibfnamefont
  {P.}~\bibnamefont {Moraes}}, \bibinfo {author} {\bibfnamefont {J.~D.}\
  \bibnamefont {Arba{\~n}il}},\ and\ \bibinfo {author} {\bibfnamefont
  {M.}~\bibnamefont {Malheiro}},\ }\href@noop {} {\bibfield  {journal}
  {\bibinfo  {journal} {Journal of Cosmology and Astroparticle Physics}\
  }\textbf {\bibinfo {volume} {2016}}\bibinfo  {number} { (06)},\ \bibinfo
  {pages} {005}}\BibitemShut {NoStop}%
\bibitem [{\citenamefont {Riley}\ \emph {et~al.}(2019)\citenamefont {Riley},
  \citenamefont {Watts}, \citenamefont {Bogdanov}, \citenamefont {Ray},
  \citenamefont {Ludlam}, \citenamefont {Guillot}, \citenamefont {Arzoumanian},
  \citenamefont {Baker}, \citenamefont {Bilous}, \citenamefont {Chakrabarty}
  \emph {et~al.}}]{riley2019nicer}%
  \BibitemOpen
\bibfield  {number} {  }\bibfield  {author} {\bibinfo {author} {\bibfnamefont
  {T.~E.}\ \bibnamefont {Riley}}, \bibinfo {author} {\bibfnamefont {A.~L.}\
  \bibnamefont {Watts}}, \bibinfo {author} {\bibfnamefont {S.}~\bibnamefont
  {Bogdanov}}, \bibinfo {author} {\bibfnamefont {P.~S.}\ \bibnamefont {Ray}},
  \bibinfo {author} {\bibfnamefont {R.~M.}\ \bibnamefont {Ludlam}}, \bibinfo
  {author} {\bibfnamefont {S.}~\bibnamefont {Guillot}}, \bibinfo {author}
  {\bibfnamefont {Z.}~\bibnamefont {Arzoumanian}}, \bibinfo {author}
  {\bibfnamefont {C.~L.}\ \bibnamefont {Baker}}, \bibinfo {author}
  {\bibfnamefont {A.~V.}\ \bibnamefont {Bilous}}, \bibinfo {author}
  {\bibfnamefont {D.}~\bibnamefont {Chakrabarty}}, \emph {et~al.},\ }\href@noop
  {} {\bibfield  {journal} {\bibinfo  {journal} {The Astrophysical Journal
  Letters}\ }\textbf {\bibinfo {volume} {887}},\ \bibinfo {pages} {L21}
  (\bibinfo {year} {2019})}\BibitemShut {NoStop}%
\bibitem [{\citenamefont {Miller}\ \emph {et~al.}(2019)\citenamefont {Miller},
  \citenamefont {Lamb}, \citenamefont {Dittmann}, \citenamefont {Bogdanov},
  \citenamefont {Arzoumanian}, \citenamefont {Gendreau}, \citenamefont
  {Guillot}, \citenamefont {Harding}, \citenamefont {Ho}, \citenamefont
  {Lattimer} \emph {et~al.}}]{miller2019psr}%
  \BibitemOpen
  \bibfield  {author} {\bibinfo {author} {\bibfnamefont {M.}~\bibnamefont
  {Miller}}, \bibinfo {author} {\bibfnamefont {F.~K.}\ \bibnamefont {Lamb}},
  \bibinfo {author} {\bibfnamefont {A.}~\bibnamefont {Dittmann}}, \bibinfo
  {author} {\bibfnamefont {S.}~\bibnamefont {Bogdanov}}, \bibinfo {author}
  {\bibfnamefont {Z.}~\bibnamefont {Arzoumanian}}, \bibinfo {author}
  {\bibfnamefont {K.~C.}\ \bibnamefont {Gendreau}}, \bibinfo {author}
  {\bibfnamefont {S.}~\bibnamefont {Guillot}}, \bibinfo {author} {\bibfnamefont
  {A.}~\bibnamefont {Harding}}, \bibinfo {author} {\bibfnamefont
  {W.}~\bibnamefont {Ho}}, \bibinfo {author} {\bibfnamefont {J.}~\bibnamefont
  {Lattimer}}, \emph {et~al.},\ }\href@noop {} {\bibfield  {journal} {\bibinfo
  {journal} {The Astrophysical Journal Letters}\ }\textbf {\bibinfo {volume}
  {887}},\ \bibinfo {pages} {L24} (\bibinfo {year} {2019})}\BibitemShut
  {NoStop}%
\bibitem [{\citenamefont {Antoniadis}\ \emph {et~al.}(2013)\citenamefont
  {Antoniadis}, \citenamefont {Freire}, \citenamefont {Wex}, \citenamefont
  {Tauris}, \citenamefont {Lynch}, \citenamefont {Van~Kerkwijk}, \citenamefont
  {Kramer}, \citenamefont {Bassa}, \citenamefont {Dhillon}, \citenamefont
  {Driebe} \emph {et~al.}}]{antoniadis2013massive}%
  \BibitemOpen
  \bibfield  {author} {\bibinfo {author} {\bibfnamefont {J.}~\bibnamefont
  {Antoniadis}}, \bibinfo {author} {\bibfnamefont {P.~C.}\ \bibnamefont
  {Freire}}, \bibinfo {author} {\bibfnamefont {N.}~\bibnamefont {Wex}},
  \bibinfo {author} {\bibfnamefont {T.~M.}\ \bibnamefont {Tauris}}, \bibinfo
  {author} {\bibfnamefont {R.~S.}\ \bibnamefont {Lynch}}, \bibinfo {author}
  {\bibfnamefont {M.~H.}\ \bibnamefont {Van~Kerkwijk}}, \bibinfo {author}
  {\bibfnamefont {M.}~\bibnamefont {Kramer}}, \bibinfo {author} {\bibfnamefont
  {C.}~\bibnamefont {Bassa}}, \bibinfo {author} {\bibfnamefont {V.~S.}\
  \bibnamefont {Dhillon}}, \bibinfo {author} {\bibfnamefont {T.}~\bibnamefont
  {Driebe}}, \emph {et~al.},\ }\href@noop {} {\bibfield  {journal} {\bibinfo
  {journal} {Science}\ }\textbf {\bibinfo {volume} {340}},\ \bibinfo {pages}
  {1233232} (\bibinfo {year} {2013})}\BibitemShut {NoStop}%
\bibitem [{\citenamefont {Fonseca}\ \emph {et~al.}(2021)\citenamefont
  {Fonseca}, \citenamefont {Cromartie}, \citenamefont {Pennucci}, \citenamefont
  {Ray}, \citenamefont {Kirichenko}, \citenamefont {Ransom}, \citenamefont
  {Demorest}, \citenamefont {Stairs}, \citenamefont {Arzoumanian},
  \citenamefont {Guillemot} \emph {et~al.}}]{fonseca2021refined}%
  \BibitemOpen
  \bibfield  {author} {\bibinfo {author} {\bibfnamefont {E.}~\bibnamefont
  {Fonseca}}, \bibinfo {author} {\bibfnamefont {H.~T.}\ \bibnamefont
  {Cromartie}}, \bibinfo {author} {\bibfnamefont {T.~T.}\ \bibnamefont
  {Pennucci}}, \bibinfo {author} {\bibfnamefont {P.~S.}\ \bibnamefont {Ray}},
  \bibinfo {author} {\bibfnamefont {A.~Y.}\ \bibnamefont {Kirichenko}},
  \bibinfo {author} {\bibfnamefont {S.~M.}\ \bibnamefont {Ransom}}, \bibinfo
  {author} {\bibfnamefont {P.~B.}\ \bibnamefont {Demorest}}, \bibinfo {author}
  {\bibfnamefont {I.~H.}\ \bibnamefont {Stairs}}, \bibinfo {author}
  {\bibfnamefont {Z.}~\bibnamefont {Arzoumanian}}, \bibinfo {author}
  {\bibfnamefont {L.}~\bibnamefont {Guillemot}}, \emph {et~al.},\ }\href@noop
  {} {\bibfield  {journal} {\bibinfo  {journal} {The Astrophysical Journal
  Letters}\ }\textbf {\bibinfo {volume} {915}},\ \bibinfo {pages} {L12}
  (\bibinfo {year} {2021})}\BibitemShut {NoStop}%
\bibitem [{\citenamefont {Cromartie}\ \emph {et~al.}(2020)\citenamefont
  {Cromartie}, \citenamefont {Fonseca}, \citenamefont {Ransom}, \citenamefont
  {Demorest}, \citenamefont {Arzoumanian}, \citenamefont {Blumer},
  \citenamefont {Brook}, \citenamefont {DeCesar}, \citenamefont {Dolch},
  \citenamefont {Ellis} \emph {et~al.}}]{cromartie2020relativistic}%
  \BibitemOpen
  \bibfield  {author} {\bibinfo {author} {\bibfnamefont {H.~T.}\ \bibnamefont
  {Cromartie}}, \bibinfo {author} {\bibfnamefont {E.}~\bibnamefont {Fonseca}},
  \bibinfo {author} {\bibfnamefont {S.~M.}\ \bibnamefont {Ransom}}, \bibinfo
  {author} {\bibfnamefont {P.~B.}\ \bibnamefont {Demorest}}, \bibinfo {author}
  {\bibfnamefont {Z.}~\bibnamefont {Arzoumanian}}, \bibinfo {author}
  {\bibfnamefont {H.}~\bibnamefont {Blumer}}, \bibinfo {author} {\bibfnamefont
  {P.~R.}\ \bibnamefont {Brook}}, \bibinfo {author} {\bibfnamefont {M.~E.}\
  \bibnamefont {DeCesar}}, \bibinfo {author} {\bibfnamefont {T.}~\bibnamefont
  {Dolch}}, \bibinfo {author} {\bibfnamefont {J.~A.}\ \bibnamefont {Ellis}},
  \emph {et~al.},\ }\href@noop {} {\bibfield  {journal} {\bibinfo  {journal}
  {Nature Astronomy}\ }\textbf {\bibinfo {volume} {4}},\ \bibinfo {pages} {72}
  (\bibinfo {year} {2020})}\BibitemShut {NoStop}%
\bibitem [{\citenamefont {Romani}\ \emph {et~al.}(2022)\citenamefont {Romani},
  \citenamefont {Kandel}, \citenamefont {Filippenko}, \citenamefont {Brink},\
  and\ \citenamefont {Zheng}}]{romani2022psr}%
  \BibitemOpen
  \bibfield  {author} {\bibinfo {author} {\bibfnamefont {R.~W.}\ \bibnamefont
  {Romani}}, \bibinfo {author} {\bibfnamefont {D.}~\bibnamefont {Kandel}},
  \bibinfo {author} {\bibfnamefont {A.~V.}\ \bibnamefont {Filippenko}},
  \bibinfo {author} {\bibfnamefont {T.~G.}\ \bibnamefont {Brink}},\ and\
  \bibinfo {author} {\bibfnamefont {W.}~\bibnamefont {Zheng}},\ }\href@noop {}
  {\bibfield  {journal} {\bibinfo  {journal} {The Astrophysical Journal
  Letters}\ }\textbf {\bibinfo {volume} {934}},\ \bibinfo {pages} {L17}
  (\bibinfo {year} {2022})}\BibitemShut {NoStop}%
\bibitem [{\citenamefont {Demorest}\ \emph {et~al.}(2010)\citenamefont
  {Demorest}, \citenamefont {Pennucci}, \citenamefont {Ransom}, \citenamefont
  {Roberts},\ and\ \citenamefont {Hessels}}]{demorest2010two}%
  \BibitemOpen
  \bibfield  {author} {\bibinfo {author} {\bibfnamefont {P.~B.}\ \bibnamefont
  {Demorest}}, \bibinfo {author} {\bibfnamefont {T.}~\bibnamefont {Pennucci}},
  \bibinfo {author} {\bibfnamefont {S.}~\bibnamefont {Ransom}}, \bibinfo
  {author} {\bibfnamefont {M.}~\bibnamefont {Roberts}},\ and\ \bibinfo {author}
  {\bibfnamefont {J.}~\bibnamefont {Hessels}},\ }\href@noop {} {\bibfield
  {journal} {\bibinfo  {journal} {nature}\ }\textbf {\bibinfo {volume} {467}},\
  \bibinfo {pages} {1081} (\bibinfo {year} {2010})}\BibitemShut {NoStop}%
\bibitem [{\citenamefont {Abbott}\ \emph {et~al.}(2018)\citenamefont {Abbott},
  \citenamefont {Abbott}, \citenamefont {Abbott}, \citenamefont {Acernese},
  \citenamefont {Ackley}, \citenamefont {Adams}, \citenamefont {Adams},
  \citenamefont {Addesso}, \citenamefont {Adhikari}, \citenamefont {Adya} \emph
  {et~al.}}]{abbott2018gw170817}%
  \BibitemOpen
  \bibfield  {author} {\bibinfo {author} {\bibfnamefont {B.~P.}\ \bibnamefont
  {Abbott}}, \bibinfo {author} {\bibfnamefont {R.}~\bibnamefont {Abbott}},
  \bibinfo {author} {\bibfnamefont {T.}~\bibnamefont {Abbott}}, \bibinfo
  {author} {\bibfnamefont {F.}~\bibnamefont {Acernese}}, \bibinfo {author}
  {\bibfnamefont {K.}~\bibnamefont {Ackley}}, \bibinfo {author} {\bibfnamefont
  {C.}~\bibnamefont {Adams}}, \bibinfo {author} {\bibfnamefont
  {T.}~\bibnamefont {Adams}}, \bibinfo {author} {\bibfnamefont
  {P.}~\bibnamefont {Addesso}}, \bibinfo {author} {\bibfnamefont {R.~X.}\
  \bibnamefont {Adhikari}}, \bibinfo {author} {\bibfnamefont {V.~B.}\
  \bibnamefont {Adya}}, \emph {et~al.},\ }\href@noop {} {\bibfield  {journal}
  {\bibinfo  {journal} {Physical review letters}\ }\textbf {\bibinfo {volume}
  {121}},\ \bibinfo {pages} {161101} (\bibinfo {year} {2018})}\BibitemShut
  {NoStop}%
\bibitem [{\citenamefont {Had{\v{z}}i{\'c}}\ and\ \citenamefont
  {Lin}(2021)}]{hadvzic2021turning}%
  \BibitemOpen
  \bibfield  {author} {\bibinfo {author} {\bibfnamefont {M.}~\bibnamefont
  {Had{\v{z}}i{\'c}}}\ and\ \bibinfo {author} {\bibfnamefont {Z.}~\bibnamefont
  {Lin}},\ }\href@noop {} {\bibfield  {journal} {\bibinfo  {journal}
  {Communications in Mathematical Physics}\ }\textbf {\bibinfo {volume}
  {387}},\ \bibinfo {pages} {729} (\bibinfo {year} {2021})}\BibitemShut
  {NoStop}%
\bibitem [{\citenamefont {Harrison}\ \emph {et~al.}(1965)\citenamefont
  {Harrison}, \citenamefont {Thorne}, \citenamefont {Wakano},\ and\
  \citenamefont {Wheeler}}]{harrison1965gravitation}%
  \BibitemOpen
  \bibfield  {author} {\bibinfo {author} {\bibfnamefont {B.~K.}\ \bibnamefont
  {Harrison}}, \bibinfo {author} {\bibfnamefont {K.~S.}\ \bibnamefont
  {Thorne}}, \bibinfo {author} {\bibfnamefont {M.}~\bibnamefont {Wakano}},\
  and\ \bibinfo {author} {\bibfnamefont {J.~A.}\ \bibnamefont {Wheeler}},\
  }\href@noop {} {\bibfield  {journal} {\bibinfo  {journal} {Gravitation Theory
  and Gravitational Collapse}\ } (\bibinfo {year} {1965})}\BibitemShut
  {NoStop}%
\bibitem [{\citenamefont {Dexheimer}\ \emph {et~al.}(2012)\citenamefont
  {Dexheimer}, \citenamefont {Negreiros},\ and\ \citenamefont
  {Schramm}}]{dexheimer2012hybrid}%
  \BibitemOpen
  \bibfield  {author} {\bibinfo {author} {\bibfnamefont {V.}~\bibnamefont
  {Dexheimer}}, \bibinfo {author} {\bibfnamefont {R.}~\bibnamefont
  {Negreiros}},\ and\ \bibinfo {author} {\bibfnamefont {S.}~\bibnamefont
  {Schramm}},\ }\href@noop {} {\bibfield  {journal} {\bibinfo  {journal} {The
  European Physical Journal A}\ }\textbf {\bibinfo {volume} {48}},\ \bibinfo
  {pages} {1} (\bibinfo {year} {2012})}\BibitemShut {NoStop}%
\bibitem [{\citenamefont {Kokkotas}\ and\ \citenamefont
  {Ruoff}(2001)}]{kokkotas2001radial}%
  \BibitemOpen
  \bibfield  {author} {\bibinfo {author} {\bibfnamefont {K.}~\bibnamefont
  {Kokkotas}}\ and\ \bibinfo {author} {\bibfnamefont {J.}~\bibnamefont
  {Ruoff}},\ }\href@noop {} {\bibfield  {journal} {\bibinfo  {journal}
  {Astronomy \& Astrophysics}\ }\textbf {\bibinfo {volume} {366}},\ \bibinfo
  {pages} {565} (\bibinfo {year} {2001})}\BibitemShut {NoStop}%
\bibitem [{\citenamefont {Bocquet}\ \emph {et~al.}(1995)\citenamefont
  {Bocquet}, \citenamefont {Bonazzola}, \citenamefont {Gourgoulhon},\ and\
  \citenamefont {Novak}}]{bocquet1995rotating}%
  \BibitemOpen
  \bibfield  {author} {\bibinfo {author} {\bibfnamefont {M.}~\bibnamefont
  {Bocquet}}, \bibinfo {author} {\bibfnamefont {S.}~\bibnamefont {Bonazzola}},
  \bibinfo {author} {\bibfnamefont {E.}~\bibnamefont {Gourgoulhon}},\ and\
  \bibinfo {author} {\bibfnamefont {J.}~\bibnamefont {Novak}},\ }\href@noop {}
  {\bibfield  {journal} {\bibinfo  {journal} {arXiv preprint gr-qc/9503044}\ }
  (\bibinfo {year} {1995})}\BibitemShut {NoStop}%
\bibitem [{\citenamefont {Chatterjee}\ \emph {et~al.}(2015)\citenamefont
  {Chatterjee}, \citenamefont {Elghozi}, \citenamefont {Novak},\ and\
  \citenamefont {Oertel}}]{chatterjee2015consistent}%
  \BibitemOpen
  \bibfield  {author} {\bibinfo {author} {\bibfnamefont {D.}~\bibnamefont
  {Chatterjee}}, \bibinfo {author} {\bibfnamefont {T.}~\bibnamefont {Elghozi}},
  \bibinfo {author} {\bibfnamefont {J.}~\bibnamefont {Novak}},\ and\ \bibinfo
  {author} {\bibfnamefont {M.}~\bibnamefont {Oertel}},\ }\href@noop {}
  {\bibfield  {journal} {\bibinfo  {journal} {Monthly Notices of the Royal
  Astronomical Society}\ }\textbf {\bibinfo {volume} {447}},\ \bibinfo {pages}
  {3785} (\bibinfo {year} {2015})}\BibitemShut {NoStop}%
\bibitem [{\citenamefont {Franzon}\ \emph {et~al.}(2016)\citenamefont
  {Franzon}, \citenamefont {Dexheimer},\ and\ \citenamefont
  {Schramm}}]{franzon2016self}%
  \BibitemOpen
  \bibfield  {author} {\bibinfo {author} {\bibfnamefont {B.}~\bibnamefont
  {Franzon}}, \bibinfo {author} {\bibfnamefont {V.}~\bibnamefont {Dexheimer}},\
  and\ \bibinfo {author} {\bibfnamefont {S.}~\bibnamefont {Schramm}},\
  }\href@noop {} {\bibfield  {journal} {\bibinfo  {journal} {Monthly Notices of
  the Royal Astronomical Society}\ }\textbf {\bibinfo {volume} {456}},\
  \bibinfo {pages} {2937} (\bibinfo {year} {2016})}\BibitemShut {NoStop}%
\bibitem [{\citenamefont {Shapiro}\ and\ \citenamefont
  {Teukolsky}(2024)}]{shapiro2024black}%
  \BibitemOpen
  \bibfield  {author} {\bibinfo {author} {\bibfnamefont {S.~L.}\ \bibnamefont
  {Shapiro}}\ and\ \bibinfo {author} {\bibfnamefont {S.~A.}\ \bibnamefont
  {Teukolsky}},\ }\href@noop {} {\emph {\bibinfo {title} {Black holes, white
  dwarfs and neutron stars: the physics of compact objects}}}\ (\bibinfo
  {publisher} {John Wiley \& Sons},\ \bibinfo {year} {2024})\BibitemShut
  {NoStop}%
\bibitem [{\citenamefont {Cardall}\ \emph {et~al.}(2001)\citenamefont
  {Cardall}, \citenamefont {Prakash},\ and\ \citenamefont
  {Lattimer}}]{cardall2001effects}%
  \BibitemOpen
  \bibfield  {author} {\bibinfo {author} {\bibfnamefont {C.~Y.}\ \bibnamefont
  {Cardall}}, \bibinfo {author} {\bibfnamefont {M.}~\bibnamefont {Prakash}},\
  and\ \bibinfo {author} {\bibfnamefont {J.~M.}\ \bibnamefont {Lattimer}},\
  }\href@noop {} {\bibfield  {journal} {\bibinfo  {journal} {The Astrophysical
  Journal}\ }\textbf {\bibinfo {volume} {554}},\ \bibinfo {pages} {322}
  (\bibinfo {year} {2001})}\BibitemShut {NoStop}%
\end{thebibliography}%

\end{document}